\documentclass[
reprint, % temporary %
english,aps,prl,superscriptaddress]{revtex4-1}
\usepackage[latin9]{inputenc}
\usepackage{color}
\usepackage{babel}
\usepackage{amsmath}
\usepackage{amssymb}
\usepackage{graphicx}
\usepackage[unicode=true,pdfusetitle,bookmarks=true,bookmarksnumbered=false,bookmarksopen=false, breaklinks=false,pdfborder={0 0 1},backref=false,colorlinks=true]{hyperref}
\hypersetup{urlcolor=blue, citecolor=blue, hyperfootnotes=blue, linkcolor=blue}
\usepackage{breakurl}

\makeatletter
\usepackage{txfonts}
\usepackage{mathrsfs}
\usepackage{epsfig}
\usepackage{amsfonts}
\usepackage[figuresright]{rotating}
\usepackage{dcolumn}
\usepackage{bm}
\usepackage{color}

\makeatletter

\begin{document}

\title{Coherent pair injection as a route towards the
\\enhancement of supersolid order in many-body bosonic models}

\author{Emmanouil Grigoriou}
\thanks{These authors contributed equally to the work}
\affiliation{Wilczek Quantum Center, School of Physics and Astronomy, Shanghai
Jiao Tong University, Shanghai 200240, China}
\author{Zhiyao Ning}
\thanks{These authors contributed equally to the work}
\affiliation{Graduate School of Science and Technology, University of Tsukuba, Tsukuba 305-8571, Japan}
\author{Hang Su}
\thanks{These authors contributed equally to the work}
\affiliation{School of Physics and Astronomy, Shanghai Jiao Tong University,
Shanghai 200240, China}
\author{Benjamin L\"ockler}
\affiliation{Max-Planck Institute for the Science of Light, Staudtstrasse 2, 91058
Erlangen, Germany}
\author{Ming Li}
\affiliation{Wilczek Quantum Center, School of Physics and Astronomy, Shanghai
Jiao Tong University, China}
\author{Yoshitomo Kamiya}
\thanks{Corresponding author.}
\affiliation{School of Physics and Astronomy, Shanghai Jiao Tong University,
Shanghai 200240, China}
\author{Carlos Navarrete-Benlloch}
\thanks{Corresponding author.}
\affiliation{Wilczek Quantum Center, School of Physics and Astronomy, Shanghai
Jiao Tong University, Shanghai 200240, China}
\affiliation{Max-Planck Institute for the Science of Light, Staudtstrasse 2, 91058
Erlangen, Germany}
\affiliation{Shanghai Research Center for Quantum Sciences, Shanghai 201315, China}
\affiliation{Departament d'\`Optica i Optometria i Ci\`encies de la Visi\'o, Universitat de Val\`encia, Dr. Moliner 50, 46100 Burjassot, Spain}

\begin{abstract}
Over the last couple of decades, quantum simulators have been probing quantum many-body physics with unprecedented levels of control. So far, the main focus has been on the access to novel observables and dynamical conditions related to condensed-matter models. However, the potential of quantum simulators goes beyond the traditional scope of condensed-matter physics: Being based on driven-dissipative quantum optical platforms, quantum simulators allow for processes that are typically not considered in condensed-matter physics. These processes can enrich in unexplored ways the phase diagram of well-established models. Taking the extended Bose-Hubbard model as the guiding example, in this work we examine the impact of coherent pair injection, a process readily available in, for example, superconducting circuit arrays. The interest behind this process is that, in contrast to the standard injection of single excitations, it can be configured to preserve the $U(1)$ symmetry underlying the model. We prove that this process favors both superfluid and density-wave order, as opposed to insulation or homogeneous states, thereby providing a novel route towards the access of lattice supersolidity.
\end{abstract}

\maketitle

\paragraph{Introduction.} 
Macroscopic quantum states capable of surviving decoherence constitute some of the most intricate phases of matter. Understanding properties such as superconductivity and superfluidity holds the promise for key technological applications. These exotic types of behavior emerge from the interplay between different microscopic processes involving many particles and are typically associated to the phenomenon of spontaneous symmetry breaking. Recently, Feynman's idea for quantum simulation in its modern incarnation \cite{Cirac12,Bloch12,Blatt12,AspuruGuzik12,Houck12,Georgescu14,Altman21} has become of central importance. In this emerging research field, many-body complexity relies on an ever increasing number of degrees of freedom, and in fact, extraordinary progress in the last decades has led to an explosion of experimental quantum platforms over which we have unprecedented levels of control. It is now possible to engineer systems that can simulate quantum models expected to exhibit a rich variety of phases otherwise difficult to observe. A number of lattice models, such as the Hubbard model \cite{Arovas22,Dutta15,Lewenstein07,Fisher89,Hubbard63}, are successful examples which can be experimentally implemented with, e.g., cold atoms in optical lattices \cite{Bloch12,Bloch08,Jaksch05}, photonic devices \cite{Carusotto13rev,Carusotto19rev,Giacobino20rev,JaqBloch20rev,Carusotto21rev}, and superconducting circuits \citep{Matti22,Matti21,Blais21,Krantz19,Gu17rev,Dalmonte15,Koch13Rev,Houck12}.

Until now, a wide range of many-body models borrowed from condensed-matter physics have been explored theoretically and experimentally in quantum simulators. In contrast, much less attention has been given to processes that do not traditionally appear in condensed-matter systems, but are available in modern simulators. Examples of these are the coherent injection of excitations or tailored dissipation \cite{Ciuti13,Ciuti14,Girvin15,Houck17,Schuster19}. In this work we focus on one process that stands out among this class: down-conversion, where a single excitation of a driving field is coherently transformed into two excitations of the system (and vice versa) \cite{Leghtas15,Touzard18,Leghtas20,WangCaiNB20,ChenNB21,Jiang21rev,Leghtas23,Leghtas23a,Leghtas23b}. In contrast to the common coherent injection of single excitations, it can be configured to preserve the symmetries that need to be spontaneously broken in order to build the macroscopic quantum coherence present in, e.g., superfluidity. Down-conversion has been a fundamental tool in quantum optics and modern technologies \cite{DrummondFizek04,SPDC18,GrynbergAspectFabreBook10,CNB-QOnotes}, and we believe that it can become a powerful tool in modern quantum simulators as well. Hence it is of paramount importance to understand emergent physics induced by this process.

As a first step towards this goal, here we address the question of which phases are favored by the presence of coherent pair injection. We focus on its action on the extended one-dimensional Bose-Hubbard model, where several types of (lattice) insulating and superfluid ground-state phases have been predicted to appear \cite{Kuhner2000,Batrouni06,Rossini2012,Batrouni2013,Kawaki17}. As a generic conclusion, we find that pair injection favors both density wave order and superfluidity, thus effectively extending the region of the phase diagram where supersolid order is expected to appear. This opens the possibility of stabilizing supersolid phases as robust steady states once dissipation is included in the model.

\paragraph{Model and main results.} To demonstrate the effects of pair injection on a many-body bosonic system, we consider the ground state of the extended soft-core Bose-Hubbard Hamiltonian in 1D with the periodic boundary condition $L+1\rightarrow 1$
\begin{align}
\hat{H}=\sum_{j=1}^{L}\Bigg[-\mu\hat{n}_{j} &+\frac{U}{2}\hat{n}_{j}(\hat{n}_{j}-1)
-\frac{\varepsilon}{2}\left(\hat{a}_{j}^{\dagger2}+\hat{a}_{j}^{2}\right)\nonumber
\\
 & +V\hat{n}_{j+1}\hat{n}_{j}-J\left(\hat{a}_{j}^{\dagger}\hat{a}_{j+1}+\hat{a}_{j+1}^{\dagger}\hat{a}_{j}\right)\Bigg], \label{eq:H}
\end{align}
where the bosonic operators satisfy canonical commutation relations $[\hat{a}_{j},\hat{a}_{l}]=0$ and $[\hat{a}_{j},\hat{a}_{l}^{\dagger}]=\delta_{jl}$. The operator $\hat{n}_{j}=\hat{a}_{j}^{\dagger}\hat{a}^{\;}_{j}$ is the number operator, $\mu$ is the chemical potential, $U$ ($V$) is the on-site (nearest-neighbor) repulsion energy, $J$ is the hopping rate, and $\varepsilon$ is the coherent pair injection rate. All parameters are taken real and positive throughout the work. This is an effective model in the sense that the full driven-dissipative problem will involve relaxation towards steady states that inherit their properties from this Hamiltonian (see \cite{SupMat} and the conclusions for more details). It is also important to note that the continuous $U(1)$ symmetry characteristic of particle-conserving Bose-Hubbard problems is replaced by a discrete $Z_{2}$ symmetry $\hat{a}_{j}\rightarrow-\hat{a}_{j}$ for $\varepsilon\neq0$.

\begin{figure}[t]
\includegraphics[width=0.75\columnwidth]{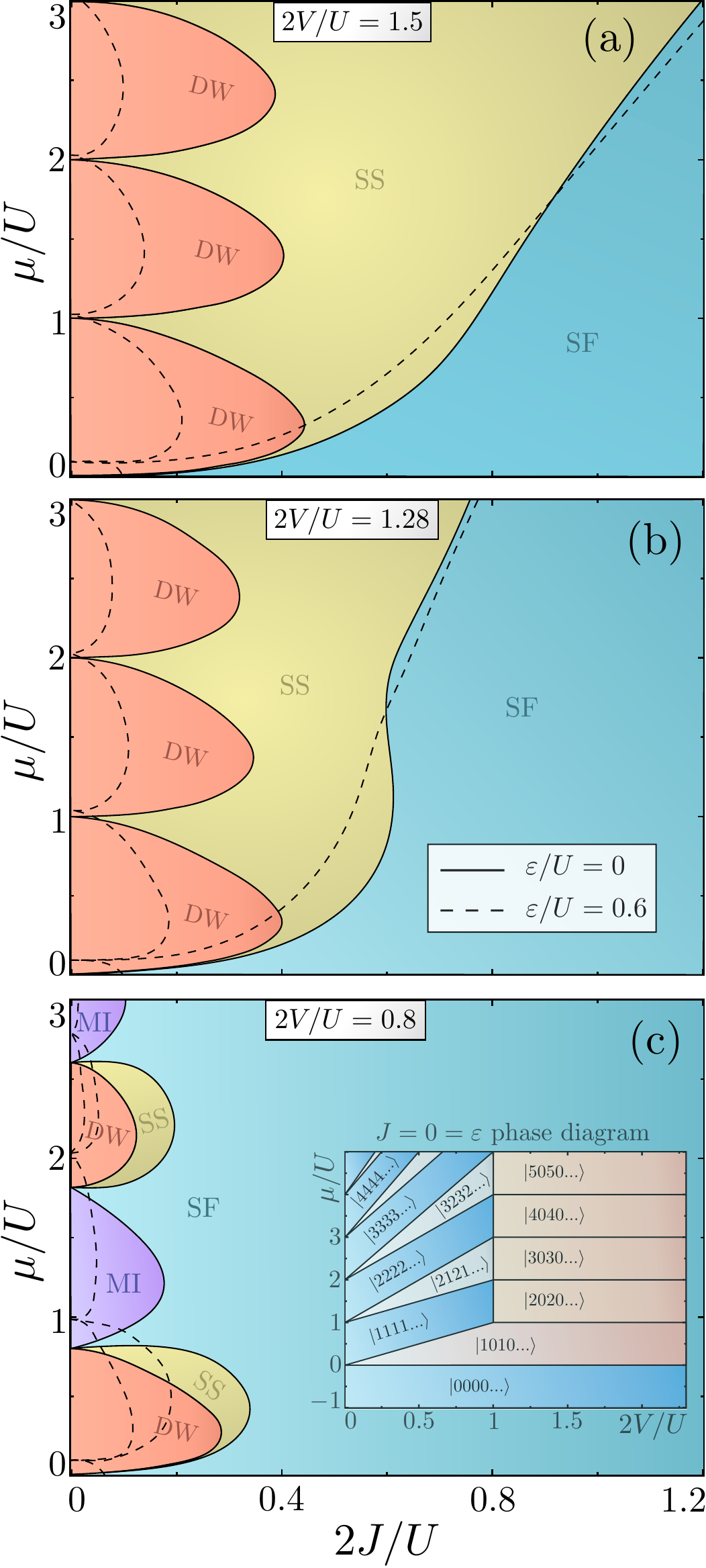}
\caption{(Color online) MF ground-state phase diagram in the $(2J/U,\mu/U)$ plane for representative values of $2V/U$. DW and MI (SF and SS) refer to density-wave and Mott insulating (homogenous-superfluid and supersolid) phases, respectively. The solid lines correspond to $\varepsilon=0$ while the dashed lines represent how the boundaries move as $\varepsilon$ increases. The phase boundaries have been artificially smoothened out for clarity; see the raw data in \cite{SupMat}. The inset in (c) shows the $J=0=\varepsilon$ phase diagram in the $(\mu/U,2V/U)$ plane, with Fock states $|n_1n_2n_3n_4...\rangle$.}\label{fig:MFphasediagram}
\end{figure}

When $\varepsilon=0,$ we recover the standard extended Bose-Hubbard Hamiltonian, which displays a global $U(1)$ symmetry under $\hat{a}_{j}\rightarrow e^{\mathrm{i}\theta}\hat{a}_{j}$ $\forall j$ for any $\theta\in\mathbb{R}$, as well as \emph{translational} symmetry, $\hat{a}_{j}\rightarrow\hat{a}_{j+d}$ $\forall j$ for any $d\in\mathbb{N}$. Setting aside for the moment subtleties associated with the 1D nature of the model, we review the  structure of the mean-field (MF) phase diagram~\cite{Kuhner2000,Yamamoto2009,Rossini2012,Suthar2020} in the $(zJ/U,\mu/U)$ plane ($z$ is the lattice coordination number; $z = 2$ in our 1D case) in Fig.~\ref{fig:MFphasediagram}, which is convenient for introducing qualitative features of the model. The MF ground-state phases can be characterized according to which of the symmetries above are broken spontaneously. The boundary between phases with lattice superfluidity, which spontaneously break the $U(1)$ symmetry, and insulating phases, which do not, has the well-known `pancake stack' structure. In the presence of nearest-neighbor interactions, both of these phases can be homogeneous or staggered, leading to four types of phases: Mott insulator (MI), preserving both symmetries; density waves (DW), breaking \emph{translational} symmetry only; superfluid (SF), breaking
only $U(1)$; supersolid (SS), breaking both symmetries. For $0<2V/U<1$ the insulating domes alternate between MI and DW, with SS phases appearing as a small region over DW domes. In contrast, for $2V/U>1$ MI phases disappear, and SS phases are able to take over a larger portion of the superfluid region, generating a SS--SF boundary that tends to a straight line for large densities.

We identify ground-state phases by analyzing the correlation functions
\begin{subequations}\label{C}
\begin{align}
C_{\mathrm{SF}}(j,l) & =\langle\hat{a}_{j}^{\dagger}\hat{a}^{}_{l}\rangle,\label{a}
\\
C_{\mathrm{DW}}(j,l) & = \langle\delta\hat{n}_{j}\delta\hat{n}_{l}\rangle,
\end{align}
\end{subequations}
with density fluctuations $\delta\hat{n}_{j}=\hat{n}_{j}-\sum_{j=1}^{L}\langle\hat{n}_{j}\rangle/L$ \cite{Rossini2012}. In the insulating phases, $C_{\mathrm{SF}}(j,l)$ decays exponentially (or faster) with the distance $|j-l|$; in contrast, it remains finite or decays as a power-law in superfluid phases. Moreover, spatially ordered phases such as DW or SS present sub-exponential decay of $C_{\mathrm{DW}}(j,l)$.

In the presence of pair injection ($\varepsilon\neq0$), the rigorous presence of superfluid order is ruled out, but it still allows for spontaneous $Z_{2}$ symmetry breaking associated with the $\theta$-phase order. Whenever a remark on the discreteness of the phase is needed, we denote its corresponding phases by $Z_{2}$SF and $Z_{2}$SS. Nevertheless, we show in \cite{SupMat} that the model (\ref{eq:H}) can be extended to restore a full $U(1)$ symmetry without impacting our main conclusions.

In the following we analyze the ground states of (\ref{eq:H}) using complementary approaches: density matrix renormalization group (DMRG), a powerful variational optimization algorithm for low-dimensional systems, to demonstrate our main point conclusively; MF approximation for a more exhaustive check of the parameter space; coherent-state ansatz to obtain analytical insight; and finally Gaussian-state ansatz to confirm the robustness of the coherent-state ansatz predictions. All these methods converge to two basic conclusions. First, pair injection $\varepsilon$ generally favors the $Z_{2}$ superfluid phases over the insulating ones; in other words, the insulating regions of the phase diagram shrink as we increase $\varepsilon$. Second, more intriguingly, deep in the region with broken phase symmetry, $\varepsilon$ favors $Z_{2}$SS order over $Z_{2}$SF order. 

\begin{figure}
\includegraphics[width=\hsize]{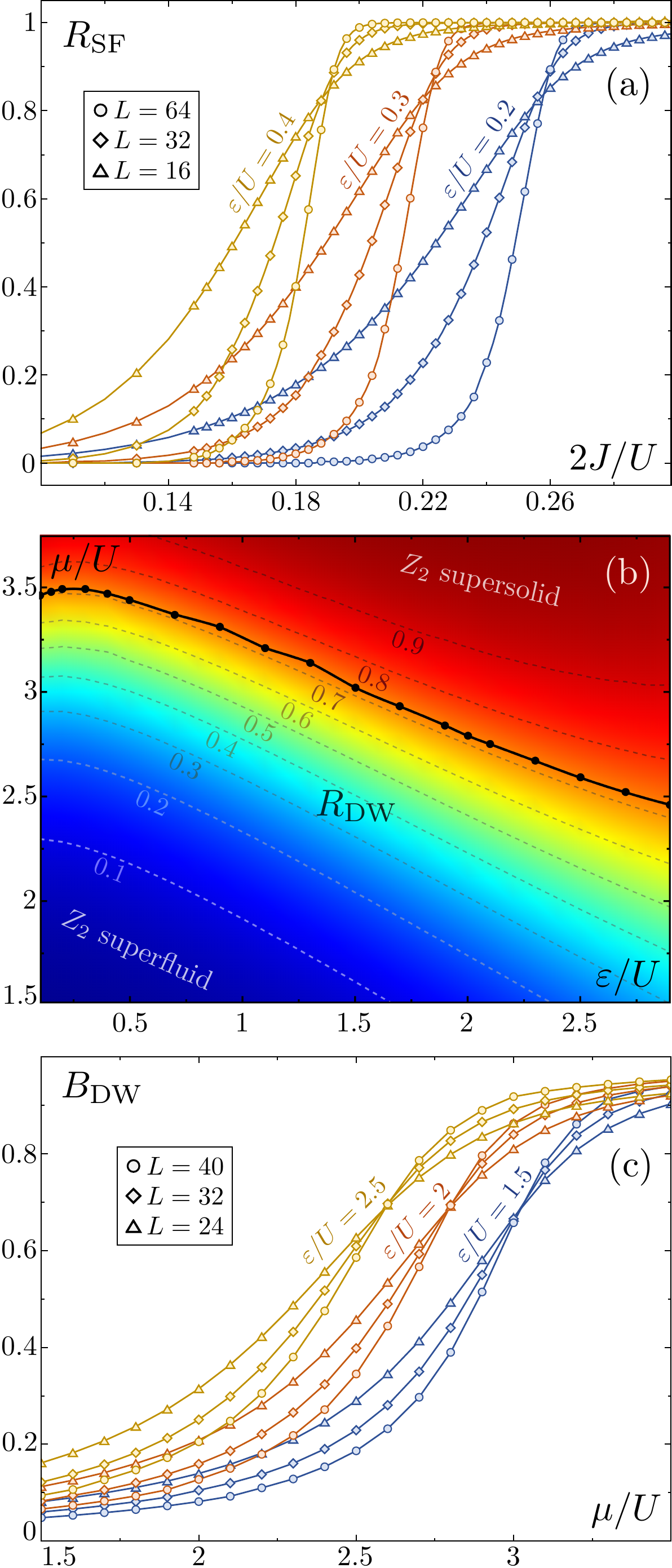}
\caption{(Color online) (a) $R_{\text{SF}}$ as a function of $2J/U$ for different values of $L$ and $\varepsilon/U$ with $\mu/U=1.8$ and $2V/U = 1.5$. As $\varepsilon/U$ increases, a smaller value of $2J/U$ can bring the system to a quantum critical point separating insulating and $Z_{2}$ superfluid phases. (b) $R_{\text{DW}}$ in the $(\mu/U,\varepsilon/U)$ plane with $2J/U=0.8$ for $L=16$. The $Z_2$ phase symmetry is broken in the entire region shown here. (c) Binder parameter $B_\mathrm{DW}$ as a function of $\mu/U$ for different values of $L$ and $\varepsilon/U$ with $2J/U=0.8$. The crossing points determine the critical $\mu/U$ for each $\varepsilon/U$ and are shown in (b) as black dots. The latest two plots show that $\varepsilon$ clearly favors density-wave order in the superfluid region, that is, supersolidity.}\label{fig:DMRG}
\end{figure}

\paragraph{DMRG results.} We run DMRG using a matrix-product-state ansatz \cite{Schollwock11,feiguin2011density,schmidt2009exact}. To simulate the soft-core bosonic model we introduce a truncation in the maximum occupation number per site, $N_{\mathrm{L}}$. The accuracy of the ansatz is controlled by the bond dimension $\chi$, which determines the degree of many-body entanglement available in the ansatz. In the results presented below, we have checked convergence of the correlation functions \eqref{C}, as well as the energy and the bipartite entanglement entropy, requiring $N_{\mathrm{L}}=10$ and $\chi=4000$ in the hardest cases.

To determine the ground state for a given parameter set, we perform a finite-size scaling of the correlation ratios~\cite{Tomita2002}
\begin{equation}
R_{\mathrm{SF}}=\frac{C_{\mathrm{SF}}(L/2)}{C_{\mathrm{SF}}(L/4)}\quad\text{and}\quad R_{\mathrm{DW}}=\frac{C_{\mathrm{DW}}(L/2)}{C_{\mathrm{DW}}(L/4)},
\label{eq:R}
\end{equation}
where $C_{\mathrm{SF/DW}}(d) \equiv C_{\mathrm{SF/DW}}(j,j+d)$ and the ratios are independent of $j$. These ratios are expected to approach $1$ as $L\rightarrow\infty$ when
the system has long-range order, or fall to $0$ if the corresponding correlation function decays exponentially. We also use
the Binder cumulant~\cite{Binder1981}
\begin{equation}
B_\mathrm{DW} = \frac{1}{2} \left(3 - \frac{\langle\hat{\phi}^{4}\rangle}{\langle\hat{\phi}^{2}\rangle^{2}} \right),
\label{Binder}
\end{equation}
which is defined to behave as $B_\mathrm{DW}\to1$ ($B_\mathrm{DW}\to0$) for $L\rightarrow\infty$ in the phase with (without) density-wave order. $\hat{\phi}=\sum_{l=1}^{L/2}(\hat{n}_{2l-1}-\hat{n}_{2l})$ is the density-wave order parameter. $B_\mathrm{DW}$ turns out to be less sensitive to finite size effects in our model and thus better suited for analyzing a density-wave transition. $B_\mathrm{DW}$, as well as $R_{\mathrm{SF}}$ and $R_{\mathrm{DW}}$, are expected to be system-size independent at a critical point and therefore form a crossing point between curves with different $L$.

In Fig.~\ref{fig:DMRG}(a), we first examine the impact of the injection rate $\varepsilon$ on the phase boundary between the insulating and $Z_{2}$ superfluid regions. We set $2V/U=1.5$ and $\mu/U=1.8$, and show $R_{\mathrm{SF}}$ as a function of $2J/U$ for different values of $\varepsilon/U$. The observed tendency towards a step function for larger $L$ suggests a second-order transition. We also observe that the superfluid phase is enlarged as $\varepsilon$ increases (the crossing occurs for lower $2J/U$). The same conclusion is drawn for any other set of parameters we have tested with DMRG.

In Figs. \ref{fig:DMRG}(b) and (c) we analyze the $Z_{2}$SF--$Z_{2}$SS boundary. We set $2V/U=1.5$ and $2J/U=0.8$, for which we confirm $R_{\mathrm{SF}}\rightarrow 1$ as $L\rightarrow\infty$ irrespective of the remaining parameters, meaning that the ground state retains the $Z_{2}$ superfluid order in the entire ($\mu/U$, $\varepsilon/U$) plane shown here. Figure~\ref{fig:DMRG}(b) shows $R_{\mathrm{DW}}$ in the $(\mu/U,\varepsilon/U)$ parameter space for $L=16$, together with the critical points (thick solid line) that we determine by performing a finite-size scaling analysis of $B_\mathrm{DW}$, as shown in Fig.~\ref{fig:DMRG}(c). Except for very low densities (small $2J/U$ and $\varepsilon/U$), it is clear that the boundary separating the $Z_{2}$SF ($R_{\mathrm{DW}}\rightarrow0$) and the $Z_{2}$SS ($R_{\mathrm{DW}}\rightarrow1$) regions moves down as $\varepsilon$ increases, concluding that pair injection extends the supersolid region. Moreover, for larger $\varepsilon$ there is a clear linear tendency of the boundary on $\varepsilon$. Indeed, our coherent-state ansatz (see below and \cite{SupMat}) predicts a critical $\mu$ given by
\begin{equation}
\mu_c=4J
(2V/U-1)^{-1}-\varepsilon,
\label{eq:muc}
\end{equation}
and existing for $2V/U>1$. Note that Eq.~\eqref{eq:muc} also predicts a linear relation between $\mu_c$ and $2J$ with slope $2/(2V/U-1)$, which is consistent with previous $\varepsilon=0$ numerical results~\cite{Batrouni2013} and with the MF picture described above. Hence, our high-precision DMRG calculation allows to univocally conclude that pair injection favors $Z_{2}$SS.

\paragraph{MF approach.} To show that the enhancement of supersolidity via pair injection can be expected for a wider parameter range, we extend the well-known \cite{Yamamoto2009,Iskin09, Sachdev11} MF approach by incorporating the $\varepsilon$ term \cite{SupMat}. By assuming a site-separable state with an alternating odd-even (o-e) site symmetry, $\lvert{\psi_\mathrm{MF}}\rangle = \bigotimes_{j=1}^{L/2}\vert\psi_{\text{o}}\rangle\otimes\vert\psi_{\text{e}}\rangle$, we turn (\ref{eq:H}) into a sum of local Hamiltonians, which for even and odd sites read
\begin{align}
\hat{h}_{m} & =-\mu\hat{n}_{m}-2J\phi_{\bar{m}}(\hat{a}_{m}^{\dagger}+\hat{a}_{m})-\frac{\varepsilon}{2}(\hat{a}_{m}^{\dagger2}+\hat{a}_{m}^{2})
\notag\\
 & \qquad+\frac{U}{2}\hat{n}_{m}(\hat{n}_{m}-1)+2V\rho_{\bar{m}}\hat{n}_{m},
 ~~m \in \{\mathrm{e},\mathrm{o}\},
\end{align}
where 
$\phi_{m}=\langle\hat{a}_{m}\rangle$,    $\rho_{m}=\langle\hat{n}_{m}\rangle$, $\bar{\mathrm{e}} \equiv \mathrm{o}$,
and $\bar{\mathrm{o}} \equiv \mathrm{e}$. Determining the ground state becomes then a nonlinear problem that we solve numerically in a self-consistent manner \cite{SupMat}. $Z_{2}$SF order is signaled by $\phi_{m}\neq 0$ for any $m$, while density-wave order is signaled by $\Delta\rho=\rho_{\text{o}}-\rho_{\text{e}}\neq0$, which allows us to distinguish all the phases of our interest.

\begin{figure}[t]
\includegraphics[width=1\columnwidth]{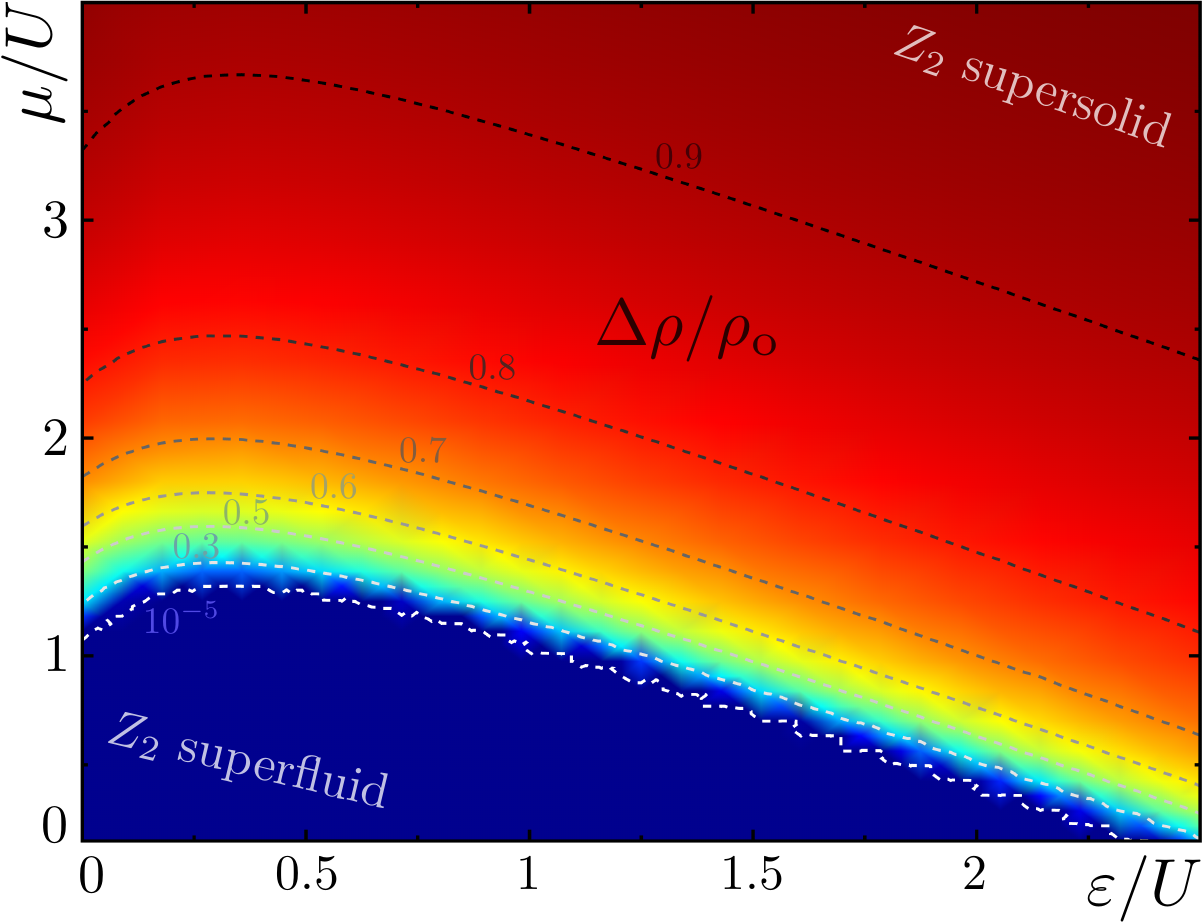}
\caption{(Color online) MF results for the DW order parameter $\Delta\rho/\rho_\text{o}=(\rho_{\text{o}}-\rho_{\text{e}})/\rho_{\text{o}}$ (odd-even relative density difference) for $2J/U = 0.8$, illustrating that pair injection in the case of a relatively small hopping rate favors $Z_2$SS over the homogeneous $Z_2$SF, except for the perturbative region (small $\varepsilon$). We assume $\rho_\text{o}\geq\rho_\text{e}$ for definiteness.}\label{fig:MFDeltaRho}
\end{figure}

The results shown in Figs. \ref{fig:MFphasediagram} and \ref{fig:MFDeltaRho} support the conclusions drawn with DMRG. The $(2J/U, \mu/U)$ phase diagram of Fig. \ref{fig:MFphasediagram} illustrates how the insulating lobes shrink as $\varepsilon$ increases. Moreover, for $2V/U>1$ the $Z_{2}$SF-$Z_{2}$SS phase boundary for large $2J/U$ asymptotically follows the coherent-state prediction~\eqref{eq:muc}, demonstrating that the supersolid phase is enlarged with $\varepsilon$. This linear tendency is most clearly observed in Fig. \ref{fig:MFDeltaRho}, where we fix $2J/U=0.8$ (for which $\phi_{m}\neq 0$) and represent $\Delta\rho/\rho_\text{o}$ in the $(\varepsilon/U,\mu/U)$ plane (for definiteness we assume $\rho_\text{o}\geq\rho_\text{e}$). The critical boundary separating $\Delta\rho=0$ ($Z_2$SF) from $\Delta\rho\neq 0$ (Z2SS) has a positive slope for small $\varepsilon/U$, but very early the slope changes sign, quickly adhering to the expected linear relation of Eq.~\eqref{eq:muc}, similarly to the DMRG results of Fig. \ref{fig:DMRG}(b). We remark that the region with initial positive slope shrinks as $2J/U$ increases.
 
\paragraph{Coherent-state ansatz.} Finally, we briefly mention on another variational treatment using a coherent-state ansatz \cite{CNB-QOnotes,WangCaiNB20,ChenNB21}, $\lvert{\psi_\mathrm{CS}}\rangle = \bigotimes_{j=1}^{L}\vert\alpha_{j}\rangle$ with $\hat{a}_{j}\vert\alpha_{j}\rangle=\alpha_{j}\vert\alpha_{j}\rangle$. For large $2J/U$, the ground state can be expected to develop a large superfluid order parameter $\phi_m$, therefore, to be dominated by a strong coherent component $\alpha_m$. Indeed, we find that the overlap between the MF state and the coherent-state ansatz is above $90\%$ at all investigated points, increasing with $2J/U$ and $\varepsilon/U$. In the coherent state ansatz, the uniform superfluid and supersolid phases are distinguished by whether the coherent amplitude is uniform ($\alpha_{j}=\alpha$, $\forall j$) or staggered ($|\alpha_{j}|\neq|\alpha_{j+1}|$ and $\alpha_{j}=\alpha_{j+2}$). The variational energy functional $\langle\hat{H}\rangle$ assumes a polynomial form as a function of the coherent amplitudes $\alpha_{j}$, which allows for deriving the formula~\eqref{eq:muc} for the $Z_2$SF--$Z_2$SS boundary \cite{SupMat}. These results are confirmed also when considering a fully general Gaussian ansatz \cite{SupMat}.

\paragraph{Conclusions and discussion.} 
As a first step towards understanding its effect on quantum simulators, we have analyzed how coherent pair injection affects the ground-state phase diagram of the extended Bose-Hubbard model. We have shown that it favors supersolid order over both insulating and homogeneous superfluid order, albeit in a model with $Z_{2}$ symmetry. Our results are however extensible to models with $U(1)$ symmetry. For example, we can consider a two-component extended Bose-Hubbard model, say, with bosonic species $a$ and $b$, and introduce pair injection $\sum_{j=1}^{L}\varepsilon(\hat{a}_{j}^{\dagger}\hat{b}_{j}^{\dagger}+\hat{a}_{j}\hat{b}_{j})$ in a flavor-selective way. This is known as non-degenerate down-conversion in quantum optics, and can be implemented in a way similar to that of the degenerate case \cite{SupMat}. The resulting model has $U(1)$ symmetry $\hat{a}_j \to e^{i\theta}\hat{a}_j$, $\hat{b}_j \to e^{-i\theta}\hat{b}_j$,  corresponding to the phase difference between the local modes. We have confirmed with a coherent ansatz that $\varepsilon$ also favors supersolid order in this model \cite{SupMat}.

Driven-dissipative quantum optical platforms have become the focus of intense research during recent decades due to their potential for realizing unexplored many-body models. Regarding our model, it is important to point out that the injection term usually comes from some sort of tunable external field that makes the injection rate time dependent~\cite{SupMat}. The time-independent Hamiltonian (\ref{eq:H}) is appears only in a picture rotating at the driving frequency. Moreover, when integrating out the degrees of freedom of the driving field, additional dissipative terms are generated that cannot be strictly described within this simple Hamiltonian formalism \cite{SupMat,Diehl16,Orus21}. Nevertheless, the interplay between coherent and dissipative processes makes the system settle into a non-equilibrium (steady) state that inherits properties present in the ground state of the rotating-picture Hamiltonian \cite{MeystreWallsBook,CNB-QOnotes}. Hence, we hope that our work, which proves that pair injection provides a controllable knob to access supersolid phases, will trigger a further effort on the stabilization of supersolidity as a steady state through a dissipative phase transition \cite{DPT2,DPT3,DPT4,DPT5,DPT1,DPT6}.

%\txtr{Moved from second section [talking about $\varepsilon$ favoring superfluidity]: In a way, this is to be expected, since coherent pair injection is equivalent to an inverted parabolic potential that favors the instabilities at the origin of phase space required for a superfluid component to grow [CarlosQuantumOpticsNotes].}

\begin{acknowledgments}
\paragraph{Acknowledgements.} 
We thank Zi Cai, Florian Marquardt, and Tao Shi for useful discussions. CNB thanks Valentina Hopekin for help with the figures and acknowledges sponsorship from the Yangyang Development Fund, as well as support from a Shanghai talent program, the Shanghai Municipal Science and Technology Major Project (Grant No. 2019SHZDZX01), and the Generalitat Valenciana through CIDEGENT project CIDEXG/2023/18. YK acknowledges the support by the NSFC (No.~12074246 and No.~U2032213) research programs.
\end{acknowledgments}

\bibliographystyle{apsrev4-1}
\bibliography{Bibliography_PairDrivenEBHM}

\onecolumngrid 
%\appendix

\newpage
\begin{center}
\textbf{\Large{}Supplemental material}{\Large\par}
\par\end{center}

We provide here important details that have been omitted in the main text for brevity. In Section \ref{Sec:DrivenDissSupMat} we carefully introduce a model for pair injection via coupling to a driven environment, discussing the relation of the final driven-dissipative model with the time-independent Hamiltonian model that we have studied in the main text. In Section \ref{Sec:AnalyticGSlimit} we provide the exact ground states of the extended Bose-Hubbard model in the $J=0$ limit. In Section \ref{sec:Appendix-Mean-Field} we provide all details about the mean-field method, including the self-consistent numerical approach as well as an analytical approach based on perturbation theory to determine the boundary between insulating and superfluid phases. Section \ref{sec:Appendix-Coherent-State} provides a detailed derivation of the theory based on a coherent-state ansatz. In Section \ref{sec:Gaussian} we consider completely general Gaussian state Ansatzes and confirm their consistency with the coherent-state predictions as we get deeper into the superfluid region. Finally, in Section \ref{sec:ModelU1} we provide an extension of the model studied in the main text in which pair injection preserves $U(1)$ symmetry and analyze it via a coherent state ansatz.

\section{Connection of our work to driven-dissipative experimental platforms}\label{Sec:DrivenDissSupMat}

\subsection{\emph{Driven-dissipative} model for pair injection and connection to our analysis}

In the main text we have studied the ground state of the extended Bose-Hubbard Hamiltonian onto which we add a coherent pair injection term $-\varepsilon(\hat{a}_{j}^{2}+\hat{a}_{j}^{\dagger2})/2$ at each lattice site $j$. As explained there, the motivation behind our study comes from the fact that such process is accessible in modern experimental quantum simulation platforms such as superconducting circuits \cite{Leghtas15,Touzard18,Leghtas20,Jiang21rev,Leghtas23,Leghtas23a,Leghtas23b}, photonic devices \cite{Carusotto13rev,Carusotto19rev,Carusotto21rev,Amo20}, and even cold atoms \cite{Law98,Law98,You02,You22,Timmermans99,Javanainen99,Drummond00}. However, our description based on adding that time-independent Hamiltonian is a simplification, not aimed at truly modeling experiments, but just at understanding the generic effect of $\varepsilon$, in particular proving that it favors supersolidity. Here we introduce a more rigorous theoretical model for pair injection as implemented in experiments and connect it to our simplified description.

Our starting point is a system consisting of a set of identical bosonic modes with characteristic energy $\omega_{0}$. In an experiment, these can correspond, for example, to the microwave modes of a superconducting circuit \cite{Matti22,Matti21,Blais21,Krantz19,Gu17rev,Dalmonte15,Koch13Rev,Houck12}, the photonic modes of an optical cavity array \cite{Carusotto19rev,Amo21}, or the Wannier modes of cold atoms trapped in optical lattices \cite{Bloch12,Bloch08,Jaksch05}. In this experimental platforms $\omega_{0}$ is usually the dominant energy scale, so that any other processes such as hopping or interactions is implemented effectively as some sort of perturbation. We thus assume that a model including all the particle-conserving terms present on the extended Bose-Hubbard Hamiltonian has been implemented in one of such platforms:
\begin{align}
\hat{H}_{0} & =\sum_{j}\left[\omega_{0}\hat{n}_{j}+\frac{U}{2}\hat{n}_{j}(\hat{n}_{j}-1)-J\left(\hat{a}_{j}^{\dagger}\hat{a}_{j+1}+\hat{a}_{j+1}^{\dagger}\hat{a}_{j}\right)+V\hat{n}_{j+1}\hat{n}_{j}\right].
\end{align}
Let us remark that at this point, $\omega_{0}$ does not have the significance of a chemical potential, and is the scale that dominates over the rest $\omega_{0}\gg J,U,V$. However, we will show shortly that an effective chemical potential on the scale of the rest of parameters appears in these systems when driven by an external field.

Let us now explain how coherent pair injection is implemented on these systems. Being a particle non-conserving process, it requires coupling the system to some sort of environment from which we can feed and extract excitations. This implies that, in general, the description of the system has to be made in terms of a mixed state $\hat{\rho}$ rather than a pure one \cite{GardinerZollerBook,CarmichaelBook,BreuerPetruccioneBook,CNB-QOnotes} (since correlations emerge with the environment), which will then be subjected to both driving and dissipation. In the next section we provide a generic description for such a system + environment model, and rigorously integrate out the environment under standard Born-Markov assumptions \cite{GardinerZollerBook,CarmichaelBook,BreuerPetruccioneBook,CNB-QOnotes} to find a dynamical equation for the state of the system alone. The resulting system dynamics is governed by a so-called master equation that reads
\begin{equation}
\partial_t\hat{\rho}=-\mathrm{i}\left[\hat{H}(t),\hat{\rho}\right]+\sum_{j}\gamma(2\hat{a}_{j}^{2}\hat{\rho}\hat{a}_{j}^{\dagger2}-\hat{a}_{j}^{\dagger2}\hat{a}_{j}^{2}\hat{\rho}-\hat{\rho}\hat{a}_{j}^{\dagger2}\hat{a}_{j}^{2}),\label{MasterEq-Appendix}
\end{equation}
with time-dependent Hamiltonian
\begin{equation}
\hat{H}(t)=\hat{H}_{0}-\sum_{j}\frac{\varepsilon}{2}\left(e^{-2\mathrm{i}\omega_{\text{d}}t}\hat{a}_{j}^{\dagger2}+e^{2\mathrm{i}\omega_{\text{d}}t}\hat{a}_{j}^{2}\right).\label{H(t)-appendix}
\end{equation}
Here $\gamma$ is real and positive and we assume the same for $\varepsilon$ for definiteness. Driving corresponds to the time-dependent term of (\ref{H(t)-appendix}), where $\varepsilon\ll\omega_{0}$ provides the rate at which an external drive at tunable frequency $2\omega_{\text{d}}$ exchanges pairs of excitations with the system through the environment; in order for this term to play a role, $|\omega_{\text{d}}-\omega_{0}|$ must not be much larger than the driving rate $\varepsilon$, as otherwise it is highly suppressed by energy conservation (rotating-wave or secular approximation \cite{CNB-QOnotes}). The coupling to the environment has another important effect, corresponding to the second term in (\ref{MasterEq-Appendix}): the incoherent loss of pairs of system excitations at rate $\gamma$, as signaled by the presence of the $\hat{a}_{j}^{2}\hat{\rho}\hat{a}_{j}^{\dagger2}$ which effect random quantum jumps removing two excitations at a time \cite{CNB-QOnotes}.

Hence, experiments deal with a driven-dissipative problem that deviates from the problem that we have analyzed in the main text in two ways. First, the Hamiltonian is time dependent. This, however, is avoided by moving to a picture rotating at frequency $\omega_{\text{d}}$. The state in the new picture is $\hat{\rho}_{\mathrm{R}}=\hat{R}^{\dagger}(t)\hat{\rho}\hat{R}(t)$, with $\hat{R}=\exp(-\mathrm{i}\omega_{\text{d}}t\hat{N})$ where $\hat{N}=\sum_{j}\hat{a}_{j}^{\dagger}\hat{a}_{j}$ is the total number operator. This state is easily shown \cite{CNB-QOnotes} to evolve
according to the same master equation (\ref{MasterEq-Appendix}), but with a time-independent Hamiltonian
\begin{align}
\hat{H}_{\text{R}} & =e^{\mathrm{i}\omega_{\text{d}}t\hat{N}}[\hat{H}(t)-\omega_{\text{d}}\hat{N}]e^{-\mathrm{i}\omega_{\text{d}}t\hat{N}}\label{HR}
\\
&=\sum_{j}\left[(\omega_{0}-\omega_{\text{d}})\hat{n}_{j}-\frac{\varepsilon}{2}\left(\hat{a}_{j}^{2}+\hat{a}_{j}^{\dagger2}\right)+\frac{U}{2}\hat{n}_{j}(\hat{n}_{j}-1)-J\left(\hat{a}_{j}^{\dagger}\hat{a}_{j+1}+\hat{a}_{j+1}^{\dagger}\hat{a}_{j}\right)+V\hat{n}_{j+1}\hat{n}_{j}\right],\nonumber 
\end{align}
where we have used $\hat{R}^{\dagger}(t)\hat{a}\hat{R}(t)=e^{-\mathrm{i}\omega_{\text{d}}t}\hat{a}$. This is precisely the Hamiltonian that we have used in the main text. Note that an effective chemical potential $\mu=\omega_{\text{d}}-\omega_{0}$ emerges in this type of driven-dissipative experiments, which can be controlled via the detuning between the driving frequency $\omega_{\text{d}}$ and the characteristic energy $\omega_{0}$ of the bosonic modes.

The presence of dissipation at rate $\gamma$ is the second way in which the experiments deviate from the situation that we have analyzed in the main text, in which we have just considered the Hamiltonian part. In general, the interplay between the Hamiltonian and dissipation will make the system settle into some kind of steady state defined by $\partial_t\hat{\rho}_\text{R}=0$; finding such state should be the ultimate goal of any theoretical approach aiming at making precise experimental predictions. Unfortunately, being a many-body driven-dissipative problem, this is an exceedingly hard task.  However, it is well known in the few-body realm of quantum optics that the properties of such steady states are typically inherited from the properties found in the Hamiltonian when dissipative processes involved are of the simple type we deal here with. For example, Hamiltonian systems possessing squeezed coherent ground states in the rotating picture lead to squeezed coherent mixed steady states when dissipation is added \cite{CNB-QOnotes,MeystreWallsBook}. Hence, understanding the effect that pair injection has on the phases of the system's rotating-picture Hamiltonian is a crucial step in determining the type of steady states that will be available via dissipative state preparation. In our case, we have shown that supersolid states are favored at the Hamiltonian level, so our work suggests that pair injection should help stabilizing supersolid phases in driven-dissipative setups. We plan on proving this statement rigorously in the near future by applying stochastic phase-space techniques \cite{Deuar21a,Deuar21b} to the master equation (\ref{MasterEq-Appendix}).

\subsection{Derivation of the master equation}

Let us here explain how the master equation (\ref{MasterEq-Appendix}) is obtained as a model that closely describes how pair injection is implemented experimentally. While the specifics of the environment and its coupling to the system might differ on each experimental platform, we can use a generic model that has the essential ingredients that are found
on all platforms. To ease the notation, let us simplify the system to a single bosonic mode with annihilation operator $\hat{a}$ and energy $\omega_{0}$ (the generalization to the full bosonic array will be easily carried later). We consider an environment consisting of a continuous set of bosonic modes with annihilation operators $\hat{b}(\omega)$, labeled by their characteristic frequency $\omega$, satisfying canonical commutation relations $[\hat{b}(\omega),\hat{b}(\omega')]=0$ and $[\hat{b}(\omega),\hat{b}^{\dagger}(\omega')]=\delta(\omega-\omega')$
\cite{CNB-QOnotes}. As specific examples, for an optical cavity this environment models the external field in empty space \cite{Drummond91,CNB17}; for a superconducting circuit, it might be the microwave field propagating in a transmission line or the modes of a lossy circuit \cite{Leghtas15}; for cold atoms trapped in optical lattices, it might correspond to the same atoms in another internal
state \cite{Law98,Law98,You02,You22,deVega08,CNB11} or to diatomic molecules \cite{Timmermans99,Javanainen99,Drummond00} trapped in a shallow potential. The Hamiltonian describing the whole system+environment is given by
\begin{equation}
\hat{H}_{\text{s+e}}=\underbrace{\omega_{0}\hat{a}^{\dagger}\hat{a}}_{\hat{H}_{\text{sys}}}+\underbrace{\int_{\omega\in\mathcal{O}(2\omega_{0})}d\omega\;\omega\hat{b}^{\dagger}(\omega)\hat{b}(\omega)}_{\hat{H}_{\text{env}}}+\underbrace{\int_{\omega\in\mathcal{O}(2\omega_{0})}d\omega\sqrt{g(\omega)}\left[\hat{a}^{\dagger2}\hat{b}(\omega)+\hat{a}^{2}\hat{b}^{\dagger}(\omega)\right]}_{\hat{H}_{\text{int}}}.\label{Hse}
\end{equation}
where we have included a coupling between the system's mode and each environmental mode with strength $g(\omega)\ll\omega_{0}$ (note that the square root is added in the definition of the interaction so that $g$ has units of energy). Any interaction aimed at implementing pair injection needs to exchange environmental excitations with pairs of system excitations, a process that is only efficient for environmental modes with energies $\omega$ around $2\omega_{0}$. We emphasize this by limiting the integration domain to $\mathcal{O}(2\omega_{0})$, understood as frequencies around $2\omega_{0}$ within a bandwidth proportional to the energy $g$ provided by the interaction. Of course, linear coupling to environmental degrees of freedom around frequencies $\omega_0$ will most likely exist as well in experiments, say a term
\begin{equation}
\int_{\omega\in\mathcal{O}(\omega_{0})}d\omega\sqrt{g(\omega)}\left[\hat{a}^{\dagger}\hat{b}(\omega)+\hat{a}\hat{b}^{\dagger}(\omega)\right].\label{HseLin}
\end{equation}
As we comment at the end of this section, would induce conventional linear damping in the system. We focus here in the quadratic coupling of (\ref{Hse}) because it is the one we use to drive the system, ultimately generating coherent pair injection.

The way in which this type of interaction is accomplished between the system and the environment varies from platform to platform. Photonic systems systems are the neatest one, since this interaction naturally occurs inside materials with a second order nonlinear polarization response to the electric field \cite{CNB-QOnotes,MeystreWallsBook,CarmichaelBook2,Drummond91} so the electromagnetic energy inside them (the polarization field times the electric field) contains a term cubic in the field, ultimately leading precisely to (\ref{Hse}). 
In this context, such interaction is referred to as down-conversion or three-wave mixing with more generality, since one excitation of the environment is converted into two excitations of the system and vice versa, involving three `waves' in the process. In other platforms the coupling between the system and the environment takes a four-wave mixing form, $\sim\hat{b}_{+}(\omega)\hat{b}_{-}(\omega)\hat{a}^{\dagger2}+\mathrm{H.c.}$, where now the environment has two modes per frequency, which satisfy canonical commutation relations $[\hat{b}_{\sigma}(\omega),\hat{b}_{\sigma'}^{\dagger}(\omega')]=\delta_{\sigma\sigma'}\delta(\omega-\omega')$ and $[\hat{b}_{\sigma}(\omega),\hat{b}_{\sigma'}(\omega)]=0$. In superconducting circuits \cite{Leghtas15,Touzard18,Leghtas20,Jiang21rev,Leghtas23,Leghtas23a,Leghtas23b} this interaction emerges when the coupling between the system's mode at frequency $\omega_{0}$ and two environmental modes close to the sidebands $\omega_{0}\pm\Delta\omega$ (with $\Delta\omega\gg g$) occurs through a Josephson junction, whose stored energy has a quartic response to the flux crossing it. In cold atoms \cite{Law98,Law98,You02,You22}, this quartic interaction comes about when, for example, the atoms trapped in the optical lattice have 0 hyperfine magnetic number, while the environmental atoms moving in a shallow trap have $\pm 1$ hyperfine magnetic number, so conservation of angular momentum forces s-wave scattering to have the quartic form introduced above. From the theoretical point of view, this quartic system-environment just complicates slightly the technical derivation that we present below, but without adding any conceptual difference. Hence we stick with the form (\ref{Hse}) to illustrate how the master equation (\ref{MasterEq-Appendix}) comes about.

Before integrating out the environment, we need to discuss its state in the absence of coupling to the system. Here is where the word ``driving'' gains a definition: this usually refers to the fact that a coherent state is generated in the environment, for example, by making it correspond to the output of a laser or a microwave generator in, respectively, optical systems or or superconducting circuits, or to a Bose-Einstein condensate in cold atoms. Coherent states are the most classical (pure) states one can build in quantum field theory, essentially leading when used as an ansatz to classical field theory onto which vacuum fluctuations are superimposed. Such coherent states are characterized by the complex amplitude $\langle\hat{b}(\omega)\rangle\equiv\beta(\omega)e^{-\mathrm{i}\omega t}$
that they produce on the environmental modes, where $|\beta(\omega)|^{2}d\omega$ is the number of excitations present in the infinitesimal interval $[\omega,\omega+d\omega]$. As Hilbert space vectors, these coherent environmental states are written as the functional \cite{CNB-QOnotes}
\begin{equation}
|\psi_{\text{env}}[\beta](t)\rangle=\underbrace{e^{\int_{\omega\in\mathcal{O}(2\omega_{0})}d\omega\left[\beta(\omega)e^{-\mathrm{i}\omega t}\hat{b}^{\dagger}(\omega)-\beta^{*}(\omega)e^{\mathrm{i}\omega t}\hat{b}(\omega)\right]}}_{\hat{D}(t)}|\text{vac}\rangle,\label{CoherentStateEnvironment}
\end{equation}
where we have defined the environmental vacuum state that satisfies $\hat{b}(\omega)|\text{vac}\rangle=0$ and the so-called displacement operator $\hat{D}(t)$. It's not difficult to see that these states satisfy the environmental Schr\"odinger equation $\mathrm{i}\partial_{t}|\psi_{\text{env}}\rangle=\hat{H}_{\text{env}}|\psi_{\text{env}}\rangle$ {[}CNBnotesQO{]}. For our purposes, we will choose $\beta(\omega)=-\beta_{\text{d}}\delta(\omega-2\omega_{\text{d}})$, which corresponds to an environment that drives the system monochromatically at frequency $2\omega_{\text{d}}$ as we will see explicitly shortly (the factor 2 in the driving frequency and the negative driving amplitude are chosen for later convenience). In the following we take $\beta_{\text{d}}$ real and positive for definiteness.

In order to integrate out the environmental degrees of freedom, it is more convenient to work in a picture where the environment is in the vacuum state. Denoting by $\hat{\rho}^{(\text{s+e})}(t)$ the state of the system+environment in the Schr\"odinger picture, we then work in a new picture where the displacement (\ref{CoherentStateEnvironment}) is removed. To this aim, we define a new state $\hat{\rho}_{D}^{(\text{s+e})}=\hat{D}^{\dagger}(t)\hat{\rho}^{(\text{\text{s+e}})}\hat{D}(t)$ that evolves according to the Hamiltonian \cite{CNB-QOnotes}
\begin{align}
\hat{H}_{D}(t) & =\hat{D}^{\dagger}(t)\hat{H}_{\text{s+e}}\hat{D}(t)-\mathrm{i}\hat{D}^{\dagger}(t)\partial_{t}\hat{D}(t)
\\
&=\hat{H}_{\text{sys}}\underbrace{-\sqrt{g(\omega_{\text{d}})}\beta_{\text{d}}\left(e^{-2\mathrm{i}\omega_{\text{d}}t}\hat{a}^{\dagger2}+e^{2\mathrm{i}\omega_{\text{d}}t}\hat{a}^{2}\right)}_{\hat{H}_{\text{d}}(t)}+\hat{H}_{\text{env}}+\hat{H}_{\text{int}},\nonumber 
\end{align}
where we have used
\begin{subequations}
\begin{align}
\hat{D}^{\dagger}(t)\hat{b} & (\omega)\hat{D}(t)=\hat{b}(\omega)+\beta(\omega)e^{-\mathrm{i}\omega t},
\\
\partial_{t}\hat{D}(t) & =\hat{D}(t)\left\{ -\mathrm{i}\int_{-\infty}^{+\infty}d\omega\;\omega\beta(\omega)e^{-\mathrm{i}\omega t}\left[\hat{b}^{\dagger}(\omega)+\frac{\beta^{*}(\omega)e^{\mathrm{i}\omega t}}{2}\right]-\text{H.c.}\right\}.
\end{align}
\end{subequations}
This picture deals with a simplified environmental state (vacuum) at the expense of working with a more complicated system Hamiltonian $\hat{H}_{\text{sys}}^{\prime}(t)=\hat{H}_{\text{sys}}+\hat{H}_{\text{d}}(t)$, but leaving the environmental Hamiltonian $\hat{H}_{\text{env}}$ and the interaction $\hat{H}_{\text{int}}$ untouched.

Since we are about to use perturbation theory as a last step, the integration of the environmental degrees of freedom is better performed in the interaction picture. This is defined by discounting all dynamics but the one generated by $\hat{H}_{\text{int}}$. We then define the unitary operator ($\mathcal{T}$ is the time-ordering symbol \cite{CNB-QOnotes})
\begin{equation}
\hat{U}(t)=\mathcal{T}\left\{ e^{-\mathrm{i}\int_{0}^{t}dt'\hat{H}_{\text{sys}}^{\prime}(t')}\right\} e^{-\mathrm{i}\hat{H}_{\text{env}}t},
\end{equation}
which transforms the environmental operators as
\begin{equation}
\hat{U}^{\dagger}(t)\hat{b}(\omega)\hat{U}(t)=e^{-\mathrm{i}\omega t}\hat{b}(\omega).
\end{equation}
The transformed state $\hat{\rho}_{\text{I}}^{(\text{s+e})}=\hat{U}^{\dagger}(t)\hat{\rho}_{D}^{(\text{s+e})}\hat{U}(t)$ evolves then according to the von Neumann equation
\begin{equation}
\mathrm{i}\partial_{t}\hat{\rho}_{\text{I}}^{(\text{s+e})}=[\hat{H}_{\text{I}}(t),\hat{\rho}_{\text{I}}^{(\text{s+e})}],\label{vnNeumannEQ}
\end{equation}
with an interaction-picture Hamiltonian
\begin{equation}
\hat{H}_{\text{I}}(t)=\hat{U}^{\dagger}(t)\underbrace{\left[\hat{H}_{D}(t)-\hat{H}_{\text{sys}}^{\prime}(t)-\hat{H}_{\text{env}}\right]}_{\hat{H}_{\text{int}}}\hat{U}(t)=\int_{\omega\in\mathcal{O}(2\omega_{0})}d\omega\sqrt{g(\omega)}\left[e^{-\mathrm{i}\omega t}\hat{a}_{\text{I}}^{\dagger2}(t)\hat{b}(\omega)+e^{\mathrm{i}\omega t}\hat{a}_{\text{I}}^{2}(t)\hat{b}^{\dagger}(\omega)\right],
\end{equation}
where we have defined the interaction-picture system operators $\hat{a}_{\text{I}}(t)=\hat{U}^{\dagger}(t)\hat{a}\hat{U}(t)$.

We now find an evolution equation for the reduced state of the system $\hat{\rho}_{\text{I}}=\text{tr}_{\text{env}}\{\hat{\rho}_{\text{I}}^{(\text{s+e})}\}$, by keeping effects on its dynamics up to second order in the interaction. To this aim, let us first consider the integral form of the von Neumann equation (\ref{vnNeumannEQ})
\begin{equation}
\hat{\rho}_{\text{I}}^{(\text{s+e})}(t)=\hat{\rho}_{\text{I}}(0)\otimes|\text{vac}\rangle\langle\text{vac}|-\mathrm{i}\int_{0}^{t}dt'[\hat{H}_{\text{I}}(t'),\hat{\rho}_{\text{I}}^{(\text{s+e})}(t')],
\end{equation}
where we have assumed that the system and the environment start in an uncorrelated state. Inserting this back on the right-hand-side of (\ref{vnNeumannEQ}) and performing the partial trace over the environmental modes, we obtain
\begin{equation}
\partial_{t}\hat{\rho}_{\text{I}}(t)=-\int_{0}^{t}d\tau\;\text{tr}_{\text{env}}\{[\hat{H}_{\text{I}}(t),[\hat{H}_{\text{I}}(t-\tau),\hat{\rho}_{\text{I}}^{(\text{s+e})}(t-\tau)]]\},
\end{equation}
where we have used $\langle\text{vac}|\hat{H}_{\text{I}}|\text{vac}\rangle=0$ and made the integration variable change $\tau=t-t'$ for later convenience. This equation is explicitly second-order on the interaction. Hence, if we want to stay at that order of approximation, we can neglect all effects on $\hat{\rho}_{\text{I}}^{(\text{s+e})}(t-\tau)$ beyond trivial order. This is known as the Born-Markov approximation. The Born approximation can be seen as a combination of two approximations: first, the correlations between the system and the environment are neglected; then, the back-action of the system onto the (comparatively larger) environment is neglected \cite{CNB-QOnotes,CarmichaelBook,BreuerPetruccioneBook}. Hence, within the Born approximation, the state on the right-hand-side is approximated by
$\hat{\rho}_{\text{I}}^{(\text{s+e})}(t-\tau)=\hat{\rho}_{\text{I}}(t-\tau)\otimes|\text{vac}\rangle\langle\text{vac}|$, which allows performing the environmental trace explicitly, obtaining
\begin{align}
\partial_{t}\hat{\rho}_{\text{I}}(t)=\int_{0}^{t}d\tau\;\Bigl\{ C(\tau) & \left[\hat{a}_{\text{I}}^{2}(t)\hat{\rho}_{\text{I}}(t-\tau)\hat{a}_{\text{I}}^{\dagger2}(t-\tau)-\hat{\rho}_{\text{I}}(t-\tau)\hat{a}_{\text{I}}^{\dagger2}(t-\tau)\hat{a}_{\text{I}}^{2}(t)\right]\label{NonMarkovianMasterEq}
\\
+ & C^{*}(\tau)\left[\hat{a}_{\text{I}}^{2}(t-\tau)\hat{\rho}_{\text{I}}(t-\tau)\hat{a}_{\text{I}}^{\dagger2}(t)-\hat{a}_{\text{I}}^{\dagger2}(t)\hat{a}_{\text{I}}^{2}(t-\tau)\hat{\rho}_{\text{I}}(t-\tau)\right]\Bigr\},\nonumber 
\end{align}
where we have defined the environmental correlation function $C(\tau)=\int_{\omega\in\mathcal{O}(2\omega_{0})}d\omega\;g(\omega)e^{\mathrm{i}\omega\tau}$. In most common situations, the couplings $g(\omega)$ vary slowly with $\omega$, so this is a function that decays quite fast with $\tau$ as we show shortly. One invokes then the Markov approximation \cite{CNB-QOnotes,CarmichaelBook,BreuerPetruccioneBook}, where it is assumed that this decay is much faster than any evolution rate of the system, so that we can set $\tau\rightarrow0$ in the state $\hat{\rho}_{\text{I}}(t-\tau)$ and the bosonic operators $\hat{a}_{\text{I}}(t-\tau)$. In order to justify this assumption, let us remember that only frequencies around $2\omega_{0}$ contribute to the physics of the problem, so that we should not make a big error by extending the integration domain to $\omega\in[-\infty,\infty]$ and setting the value of the couplings to $g(2\omega_{0})$ $\forall\omega$. This is sometimes called the `strong' Markov approximation and leads to $C(\tau)=2\pi g(2\omega_{0})\delta(\tau)$, that is, a correlation function that decays infinitely fast (infinite decay rate), allowing for contributions only at $\tau=0$. Performing the Markov approximation (in its strong form for simplicity), we turn
(\ref{NonMarkovianMasterEq}) into
\begin{equation}
\partial_{t}\hat{\rho}_{\text{I}}(t)=g(2\omega_{0})\pi\left[2\hat{a}_{\text{I}}^{2}(t)\hat{\rho}_{\text{I}}(t)\hat{a}_{\text{I}}^{\dagger2}(t)-\hat{a}_{\text{I}}^{\dagger2}(t)\hat{a}_{\text{I}}^{2}(t)\hat{\rho}_{\text{I}}(t)-\hat{\rho}_{\text{I}}(t)\hat{a}_{\text{I}}^{\dagger2}(t)\hat{a}_{\text{I}}^{2}(t)\right].
\end{equation}
Finally, moving back to the original Schr\"odinger picture, the system's state $\hat{\rho}(t)=\hat{U}(t)\hat{\rho}_{\text{I}}(t)\hat{U}^{\dagger}(t)$ evolves according to the master equation
\begin{equation}
\partial_t\hat{\rho}=-\mathrm{i}\left[\hat{H}(t),\hat{\rho}\right]+\gamma(2\hat{a}^{2}\hat{\rho}\hat{a}^{\dagger2}-\hat{a}^{\dagger2}\hat{a}^{2}\hat{\rho}-\hat{\rho}\hat{a}^{\dagger2}\hat{a}^{2}),\label{SingleModeMasterEq-Appendix}
\end{equation}
with Hamiltonian
\begin{equation}
\hat{H}(t)=\omega_{0}\hat{a}^{\dagger}\hat{a}-\frac{\varepsilon}{2}\left(e^{-2\mathrm{i}\omega_{\text{d}}t}\hat{a}^{\dagger2}+e^{2\mathrm{i}\omega_{\text{d}}t}\hat{a}^{2}\right),\label{SingleMode-H(t)-appendix}
\end{equation}
where we made the identifications $\varepsilon=\beta_{\text{d}}g(2\omega_{\text{d}})$ and $\gamma=\pi g(2\omega_{0})$. Note that the Markov approximation requires that these parameters must be much smaller than the decay rate of the environmental correlator. This is a condition that one has to be careful to satisfy in experiments unless specifically looking for non-Markovian effects.

The master equation (\ref{SingleModeMasterEq-Appendix}) is the single-mode equivalent of the one we were set to derive, Eq. (\ref{MasterEq-Appendix}). The generalization to a bosonic array is straightforward, assuming that each mode of the array is connected to its own environment and that the extended Bose-Hubbard Hamiltonian (\ref{HR}) satisfies the Markov conditions (meaning that, in addition to $\varepsilon$ and $\gamma$, the rest of parameters $J$, $U$, $V$, and $\mu$ are also much smaller than the decay rate of the environmental correlator).

Finally, let us comment on the effect that a linear coupling of the system to the environment of the type (\ref{HseLin}) would have. Following the same procedure derived above for the quadratic coupling, we would obtain an extra dissipative term on (\ref{SingleModeMasterEq-Appendix}) of the type $\kappa(2\hat{a}\hat{\rho}\hat{a}^\dagger-\hat{a}^\dagger\hat{a}\hat{\rho}-\hat{\rho}\hat{a}^\dagger\hat{a})$, with $\kappa=\pi g(\omega_0)$

\section{Ground state in the $J=0=\varepsilon$ limit}\label{Sec:AnalyticGSlimit}

\begin{figure}[b]
\includegraphics[width=0.4\textwidth]{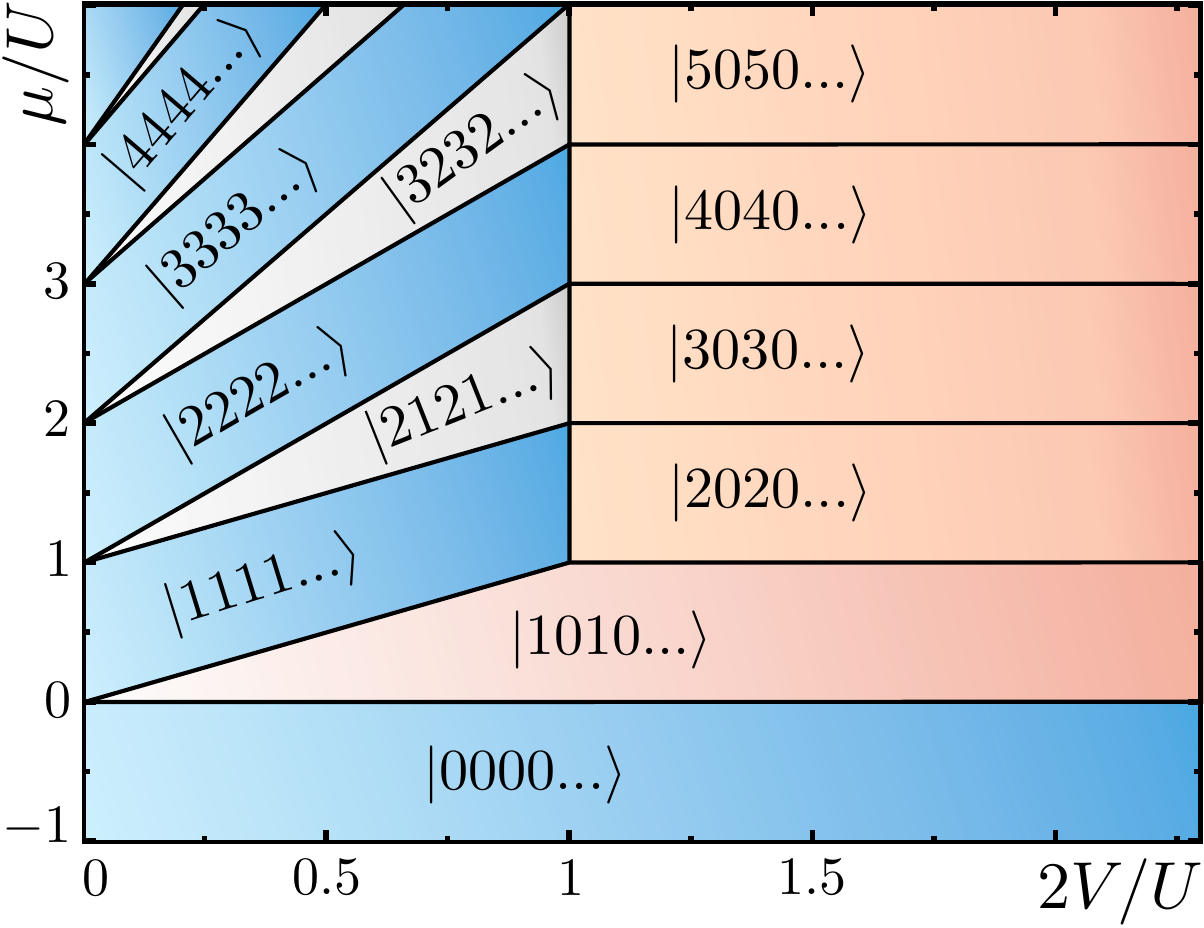} \caption{Phase diagram of the $J=0=\varepsilon$ model (extended Bose-Hubbard model at zero hopping). The labels of the states refer to the Fock numbers on each lattice site, that is, $\bigotimes_{j=1}^{L/2}|\bar{n}_{\text{o}}\rangle\otimes|\bar{n}_{\text{e}}\rangle=|n_1n_2n_3...\rangle$. as given by (\ref{n_GS}).} \label{fig:ExactPhaseDiagramZeroJ}
\end{figure}

Our work focuses on the analysis of the ground-state properties of the Hamiltonian
\begin{equation}
\hat{H}=\sum_{j}\left[-\mu\hat{n}_{j}-\frac{\varepsilon}{2}\left(\hat{a}_{j}^{2}+\hat{a}_{j}^{\dagger2}\right)+\frac{U}{2}\hat{n}_{j}(\hat{n}_{j}-1)-J\left(\hat{a}_{j}^{\dagger}\hat{a}_{j+1}+\hat{a}_{j+1}^{\dagger}\hat{a}_{j}\right)+V\hat{n}_{j+1}\hat{n}_{j}\right].\label{H_SM}
\end{equation}
While no analytic solution exists for this problem in all parameter space, it is interesting to consider the $J=0=\varepsilon$ limit, where the problem is readily solvable. In this limit, the Hamiltonian depends solely on number operators $\hat{n}_{j}$, so it is diagonalized in the Fock basis $\{\bigotimes_{j=1}^{L}|n_{j}\rangle\}_{n_{j}=0,1,...,N}$, with $\hat{n}_{j}|n\rangle=n|n\rangle$. It is not difficult to show that the ground state is given by alternating Fock numbers, particularly
$\bigotimes_{j=1}^{L/2}|\bar{n}_{\text{o}}\rangle\otimes|\bar{n}_{\text{e}}\rangle$ with
\begin{subequations}\label{n_GS}
\begin{align}
\bar{n}_{\text{o}} & =\left\{ \begin{array}{cc}
\left\lceil \frac{\mu}{U}\right\rceil  & \text{for }2V/U>1
\\
\left\lceil \frac{\mu}{U+2V}\right\rceil  & \text{for }2V/U<1
\end{array}\right.,
\\
\bar{n}_{\text{e}} & =\left\{ \begin{array}{cc}
0 & \text{for }2V/U>1
\\
\left\lceil \frac{\mu-2V}{U+2V}\right\rceil  & \text{for }2V/U<1
\end{array}\right.,
\end{align}
\end{subequations}
where $\left\lceil x\right\rceil $ is the ceiling function, which corresponds to the smallest integer not smaller than $x$. Of course, owed to the translational invariance of the model, the state $\bigotimes_{j=1}^{L/2}|\bar{n}_{\text{e}}\rangle\otimes|\bar{n}_{\text{o}}\rangle$ has the same energy, but we take the density of odd sites larger than that of even sites for definiteness. We represent the ground state in the $(2V/U,\mu/U)$ phase diagram in Fig. \ref{fig:ExactPhaseDiagramZeroJ}.

\newpage

\section{Mean Field approximation}\label{sec:Appendix-Mean-Field}

In this section we explain in detail the mean field (MF) approach. We use a numerical method (self-consistent Hartree-Fock) to determine the MF phase diagram, but also explain how to analytically obtain the insulator/superfluid boundaries in the $\varepsilon=0$ case, which serve to benchmark our numerics.

\subsection{Numerical mean field: self-consistent Hartree-Fock method}

\begin{figure}[t]
\includegraphics[width=0.8\textwidth]{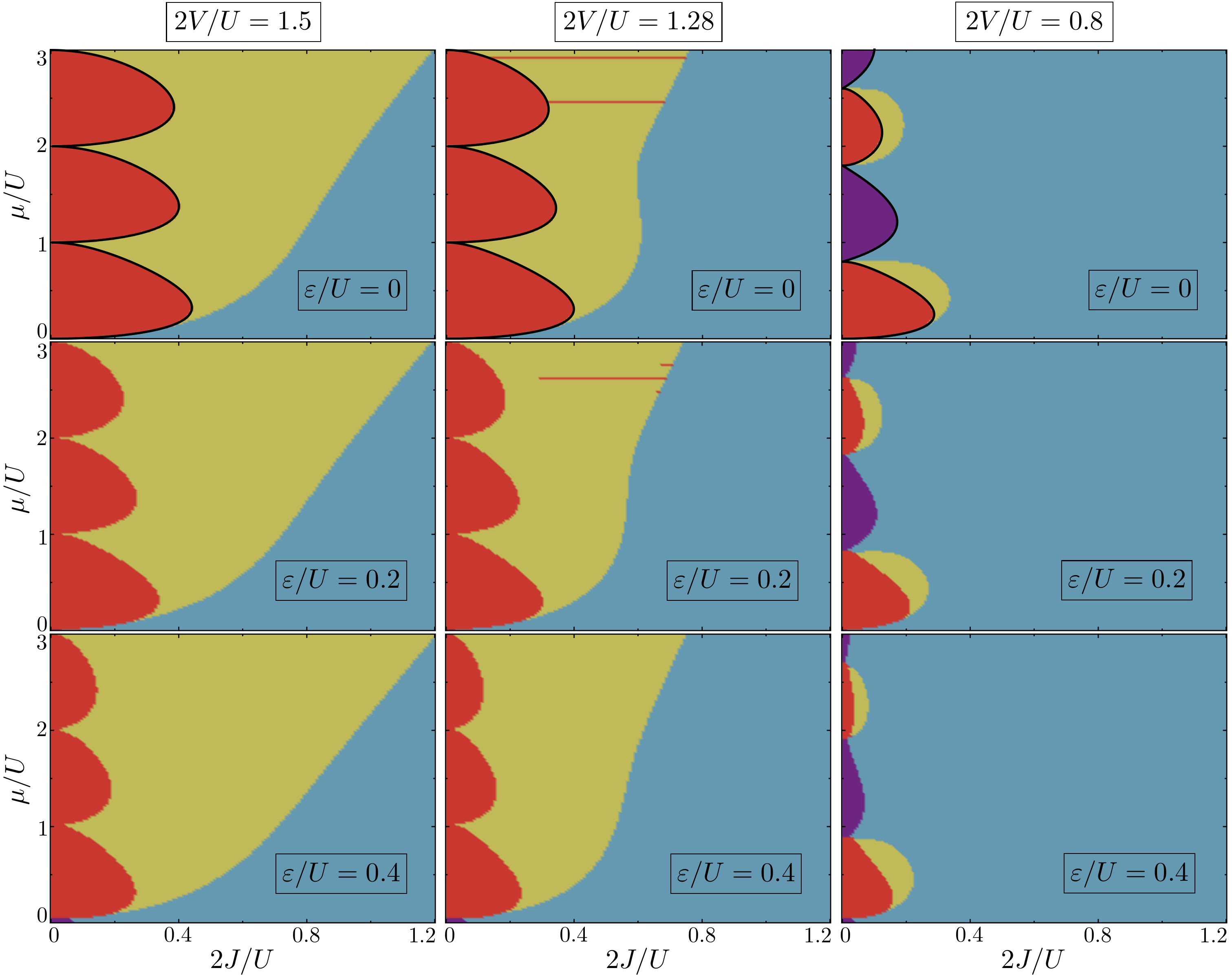} \caption{Numerical MF phase diagrams (color code: blue = SF, yellow = SS, red = DW, purple = MI). The thick black line in the first row corresponds to the analytical perturbative expression for the insulating/superfluid boundary, Eq. (\ref{InsuSuperfluidBoundaryAnalytic}).} \label{fig:MF_SM}
\end{figure}

The MF approximation is based on a separable ansatz of the type $\bigotimes_{j=1}^{L}|\psi_{j}\rangle$, which ensures that the coupling between the quantum fluctuations at different lattice sites is negligible, so that given two operators $\hat{A}$ and $\hat{B}$ acting on different sites, we have
\begin{equation}
\hat{A}\hat{B}=(\langle\hat{A}\rangle+\delta\hat{A})(\langle\hat{B}\rangle+\delta\hat{B})\approx\langle\hat{A}\rangle\langle\hat{B}\rangle+\langle\hat{A}\rangle\delta\hat{B}+\langle\hat{B}\rangle\delta\hat{A}=\langle\hat{A}\rangle\hat{B}+\langle\hat{B}\rangle\hat{A}-\langle\hat{A}\rangle\langle\hat{B}\rangle,
\end{equation}
where we have defined the fluctuation operator $\delta\hat{A}=\hat{A}-\langle\hat{A}\rangle$ and similarly for $\hat{B}$. In our Hamiltonian (\ref{H_SM}) we find two of such type of terms, the hopping and the nearest-neighbors interaction, which we rewrite under the MF approximation as
\begin{subequations}
\begin{align}
\hat{a}_{j}^{\dagger}\hat{a}_{j+1} & \approx\phi_{j}^{*}\hat{a}_{j+1}+\hat{a}_{j}^{\dagger}\phi_{j+1}-\phi_{j}^{*}\phi_{j+1},\;\;\;\text{with }\phi_{j}=\langle\hat{a}_{j}\rangle,
\\
\hat{n}_{j}\hat{n}_{j+1} & \approx\rho_{j}\hat{n}_{j+1}+\hat{n}_{j}\rho_{j+1}-\rho_{j}\rho_{j+1},\;\;\;\text{with }\rho_{j}=\langle\hat{n}_{j}\rangle.
\end{align}
\end{subequations}
The Hamiltonian takes then the form
\begin{equation}
\hat{H}\approx\sum_{j=1}^{L}\hat{h}_{j}+\mathcal{E}\equiv\hat{H}_{\text{MF}},\label{HMF}
\end{equation}
with local Hamiltonian operators
\begin{equation}
\hat{h}_{j}=-\mu_{j}\hat{n}_{j}-\frac{\varepsilon}{2}\left(\hat{a}_{j}^{2}+\hat{a}_{j}^{\dagger2}\right)+\frac{U}{2}\hat{n}_{j}(\hat{n}_{j}-1)-J[(\phi_{j-1}^{*}+\phi_{j+1}^{*})\hat{a}_{j}+\text{H.c.}],
\end{equation}
where $\mu_{j}=\mu-V(\rho_{j-1}+\rho_{j+1})$, and a constant term
\begin{equation}
\mathcal{E}=\sum_{j=1}^{L}[J(\phi_{j}^{*}\phi_{j+1}+\phi_{j}\phi_{j+1}^{*})-V\rho_{j}\rho_{j+1}].
\end{equation}
The Hamiltonian has then turned into a collection of local Hamiltonians coupled through the $\{\phi_{j},\rho_{j}\}_{j=1,2,...,L}$ expectation values. Hence, finding the ground state becomes a nonlinear problem, since the $\hat{h}_{j}$'s depend on the state one is trying to determine. The problem is simplified even further by noting that we expect the ground state to be invariant under translations by an even number of sites. Denoting by $j=\text{e}$ and $j=\text{o}$ some reference odd and even sites, respectively, so that the ansatz can be rewritten as $\bigotimes_{j=1}^{L/2}|\psi_{\text{o}}\rangle\otimes|\psi_{\text{e}}\rangle$, finding the MF ground state becomes then equivalent to finding the common ground states $|\psi_{\text{o}}\rangle$ and $|\psi_{\text{e}}\rangle$ of, respectively, the local operators
\begin{subequations}\label{hoe}
\begin{align}
\hat{h}_{\text{o}}(|\psi_{\text{e}}\rangle) & =-(\mu-2V\rho_{\text{e}})\hat{n}_{\text{o}}-\frac{\varepsilon}{2}\left(\hat{a}_{\text{o}}^{2}+\hat{a}_{\text{o}}^{\dagger2}\right)+\frac{U}{2}\hat{n}_{\text{o}}(\hat{n}_{\text{o}}-1)-2J(\phi_{\text{e}}^{*}\hat{a}_{\text{o}}+\phi_{\text{e}}\hat{a}_{\text{o}}^{\dagger}),
\\
\hat{h}_{\text{e}}(|\psi_{\text{o}}\rangle) & =-(\mu-2V\rho_{\text{o}})\hat{n}_{\text{e}}-\frac{\varepsilon}{2}\left(\hat{a}_{\text{e}}^{2}+\hat{a}_{\text{e}}^{\dagger2}\right)+\frac{U}{2}\hat{n}_{\text{e}}(\hat{n}_{\text{e}}-1)-2J(\phi_{\text{o}}^{*}\hat{a}_{\text{e}}+\phi_{\text{o}}\hat{a}_{\text{e}}^{\dagger}).
\end{align}
\end{subequations}
The notation emphasizes that
$\hat{h}_{\text{o}}$ and $\hat{h}_{\text{e}}$ depend, respectively, on the state of the even and odd sites, $|\psi_{\text{e}}\rangle$ and $|\psi_{\text{o}}\rangle$. Starting from an initial guess for these states, or rather the expectation values $\{\phi_{j},\rho_{j}\}_{j=\text{o},\text{e}}$, we find the ground state of $\hat{h}_{\text{o}}(|\psi_{\text{e}}\rangle)$ and $\hat{h}_{\text{e}}(|\psi_{\text{o}}\rangle)$, update the states $|\psi_{\text{e}}\rangle$ and $|\psi_{\text{o}}\rangle$, and iterate the procedure as many times as required for $\langle\hat{H}\rangle$ to converge. This is sometimes called a self-consistent Hartree-Fock procedure. We perform the calculation in the local (truncated) Fock basis $\{|n\rangle\}_{n=0,1,...,N}$, with $\hat{n}_{j}|n\rangle=n|n\rangle$, where $N$ is a suitable truncation ($N=20$ is usually enough to reach large accuracy in the parameter region analyzed in this work). For a given parameter set, we perform the self-consistent Hartree-Fock procedure starting from several educated or random guesses $\{\phi_{j},\rho_{j}\}_{j=\text{o},\text{e}}$, and then choose among all final states the one leading to the smallest $\langle\hat{H}\rangle$. The phase of the system is then characterized by the following conditions
\begin{equation}
\begin{array}{rl}
\text{Mott insulator (MI)}: & \phi_{\text{o}}=0=\phi_{\text{e}},\;\;\rho_{\text{o}}=\rho_{\text{e}},
\\
\text{Density-wave insulator (DW)}: & \phi_{\text{o}}=0=\phi_{\text{o}},\;\;\rho_{\text{o}}\ne\rho_{\text{e}},
\\
\text{Homogeneous }Z_{2}\text{-superfluid (SF)}: & \phi_{\text{o}}=\phi_{\text{e}}\ne0,\;\;\rho_{\text{o}}=\rho_{\text{e}},
\\
Z_{2}\text{-supersolid (SS)}: & \phi_{\text{o}}\ne\phi_{\text{e}},\;\;\rho_{\text{o}}\ne\rho_{\text{e}}.
\end{array}
\end{equation}
This is the procedure that we have used to produce the phase diagrams of Fig. 1 in the main text. Note that the boundaries of those diagrams have been smoothened for ease of presentation. For completeness, in Fig. \ref{fig:MF_SM} of this section we provide the same diagrams plotted from the raw data without any smoothing.

\subsection{Analytic MF theory for the insulating/superfluid boundary}

For $\varepsilon=0$ it is possible to find the boundaries between insulating (MI or DW) and superfluid (SF or SS) phases analytically. In short, the idea is to interpret the MF approximation to $E=\langle\hat{H}\rangle$ as sort of a Landau potential that depends on the complex parameters $\boldsymbol{\phi}=(\phi_{\text{o}},\phi_{\text{e}})^{\mathsf{T}}$ that we collect into a column vector for later convenience. In the
insulating phases, the potential has a minimum at $\boldsymbol{\phi}=0$, but this minimum flattens out as the boundary towards the superfluid phases is approached, until it becomes unstable once we cross it and new minima emerge at $\boldsymbol{\phi}\neq0$. Hence, in this section we let the parameters $\boldsymbol{\phi}$ of the MF Hamiltonian vary freely, using them as Landau `order parameters' and seeking the points of the phase diagram at which the curvature of $E(\boldsymbol{\phi})$ vanishes. This will provide analytic expressions for the insulating/superfluid boundary. Note that throughout this section the other MF parameters keep their $\rho_{j}=\langle\hat{n}_{j}\rangle$ definition. Let us now elaborate on the details of this procedure.

Given the MF variational ansatz $\bigotimes_{j=1}^{L/2}|\psi_{\text{o}}\rangle\otimes|\psi_{\text{e}}\rangle$,
we begin by defining the variational energy per pair of sites associated to the Hamiltonian (\ref{H_SM})
\begin{equation}
e=\frac{\langle\hat{H}\rangle}{L/2}=-\mu(\langle\hat{n}_{\text{o}}\rangle+\langle\hat{n}_{\text{e}}\rangle)+\frac{U}{2}(\langle\hat{n}_{\text{o}}(\hat{n}_{\text{o}}-1)\rangle+\langle\hat{n}_{\text{e}}(\hat{n}_{\text{e}}-1)\rangle)-2J\left(\langle\hat{a}_{\text{o}}^{\dagger}\rangle\langle\hat{a}_{\text{e}}\rangle+\langle\hat{a}_{\text{e}}^{\dagger}\rangle\langle\hat{a}_{\text{o}}\rangle\right)+2V\langle\hat{n}_{\text{o}}\rangle\langle\hat{n}_{\text{e}}\rangle.
\end{equation}
Similarly, writing the MF Hamiltonian per pair of sites from (\ref{HMF}) and
(\ref{hoe}) as
\begin{align}\label{hMF}
\hat{h}_{\text{MF}}\equiv\frac{\hat{H}_{\text{MF}}}{L/2}= & -\mu(\hat{n}_{\text{o}}+\hat{n}_{\text{e}})+\frac{U}{2}[\hat{n}_{\text{o}}(\hat{n}_{\text{o}}-1)+\hat{n}_{\text{e}}(\hat{n}_{\text{e}}-1)]
\\
& +2V(\langle\hat{n}_{\text{e}}\rangle\hat{n}_{\text{o}}+\langle\hat{n}_{\text{o}}\rangle\hat{n}_{\text{e}}-\langle\hat{n}_{\text{o}}\rangle\langle\hat{n}_{\text{e}}\rangle)\nonumber
\\
& -2J(\phi_{\text{e}}^{*}\hat{a}_{\text{o}}+\phi_{\text{e}}\hat{a}_{\text{o}}^{\dagger}+\phi_{\text{o}}^{*}\hat{a}_{\text{e}}+\phi_{\text{o}}\hat{a}_{\text{e}}^{\dagger}-\phi_{\text{o}}^{*}\phi_{\text{e}}-\phi_{\text{e}}^{*}\phi_{\text{o}}),\nonumber 
\end{align}
we can define the MF energy per pair of sites as
\begin{align}
e_{\text{MF}}\equiv & \langle\hat{h}_{\text{MF}}\rangle\label{eMF}
\\
= & -\mu(\langle\hat{n}_{\text{o}}\rangle+\langle\hat{n}_{\text{e}}\rangle)+\frac{U}{2}(\langle\hat{n}_{\text{o}}(\hat{n}_{\text{o}}-1)\rangle+\langle\hat{n}_{\text{e}}(\hat{n}_{\text{e}}-1)\rangle)+2V\langle\hat{n}_{\text{o}}\rangle\langle\hat{n}_{\text{e}}\rangle\nonumber 
\\
& -2J(\phi_{\text{e}}^{*}\langle\hat{a}_{\text{o}}\rangle+\phi_{\text{e}}\langle\hat{a}_{\text{o}}^{\dagger}\rangle+\phi_{\text{o}}^{*}\langle\hat{a}_{\text{e}}\rangle+\phi_{\text{o}}\langle\hat{a}_{\text{e}}^{\dagger}\rangle-\phi_{\text{o}}^{*}\phi_{\text{e}}-\phi_{\text{e}}^{*}\phi_{\text{o}}).\nonumber 
\end{align}
We then have the following relation between the variational energy and the MF energy
\begin{align}
e & =e_{\text{MF}}-2J[(\langle\hat{a}_{\text{o}}\rangle^*-\phi_{\text{o}}^{*})(\langle\hat{a}_{\text{e}}\rangle-\phi_{\text{e}})+\text{c.c.}].\label{e_gen}
\end{align}

The goal is to relate $e$ to $\boldsymbol{\phi}$ and then check its curvature at $\boldsymbol{\phi}=0$, which is built from the second-order derivatives $\mathcal{K}_{jk}=\left.\partial^{2}e/\partial\phi_{j}\partial\phi_{k}\right|_{\boldsymbol{\phi}=0}$. Hence, we don't care about the full dependence of $e$ on $\boldsymbol{\phi}$, but just about its quadratic expansion around $\boldsymbol{\phi}=0$. This is consistent with performing second-order perturbation theory in $J$ on $\hat{h}_{\text{MF}}$, which indeed leads to a quadratic form in $\boldsymbol{\phi}$ as we show next. Moreover, remember from Section \ref{Sec:AnalyticGSlimit} that for $J=0=\varepsilon$ we know the exact eigenstates of $\hat{H}$, and they have $\langle\hat{a}_{j}\rangle=0$, so $J$ is precisely the parameter that brings in the coherence $\boldsymbol{\phi}\neq0$. However, applying perturbation theory to $\hat{h}_{\text{MF}}$ is not as straightforward as it might seem, since it is a function of $\langle\hat{n}_{j}\rangle$, so it defines a nonlinear eigenproblem, where the operator we want to diagonalize depends on its eigenstates. We carefully perform such perturbation theory in the next section, but here we just provide the final result for the ground state energy, which reads
\begin{equation}
e_{\text{MF}}(\boldsymbol{\phi})\approx e_{\bar{n}_{\text{o}}\bar{n}_{\text{e}}}^{(0)}+2J(\phi_{\text{o}}^{*}\phi_{\text{e}}+\phi_{\text{o}}\phi_{\text{e}}^{*})-\frac{4J^{2}}{U}\chi_{\text{o}}|\phi_{\text{e}}|^{2}-\frac{4J^{2}}{U}\chi_{\text{e}}|\phi_{\text{o}}|^{2},\label{eMFapprox}
\end{equation}
where
\begin{align}
e_{\bar{n}_{\text{o}}\bar{n}_{\text{e}}}^{(0)} & =-\mu(\bar{n}_{\text{o}}+\bar{n}_{\text{e}})+2V\bar{n}_{\text{o}}\bar{n}_{\text{e}}+\frac{U}{2}[\bar{n}_{\text{o}}(\bar{n}_{\text{o}}-1)+\bar{n}_{\text{e}}(\bar{n}_{\text{e}}-1)],
\end{align}
is the ground-state energy of the system in the $J=0$ limit, with $\{\bar{n}_{j}\}_{j=\text{o},\text{e}}$ given by (\ref{n_GS}), and
\begin{subequations}
\begin{align}
\chi_{\text{o}} & =\frac{\bar{n}_{\text{o}}}{\mu/U-2\bar{n}_{\text{e}}V/U+1-\bar{n}_{\text{o}}}+\frac{\bar{n}_{\text{o}}+1}{\bar{n}_{\text{o}}-\mu/U+2\bar{n}_{\text{e}}V/U},
\\
\chi_{\text{e}} & =\frac{n_{\text{e}}}{\mu/U-2\bar{n}_{\text{o}}V/U+1-\bar{n}_{\text{e}}}+\frac{\bar{n}_{\text{e}}+1}{\bar{n}_{\text{e}}-\mu/U+2\bar{n}_{\text{o}}V/U}.
\end{align}
\end{subequations}
Note that $\chi_{j}\geq0$.

In order to turn the variational energy per pair of sites $e$ of Eq. (\ref{e_gen}) into a function of $\boldsymbol{\phi}$, we also have to relate the expectation values $\langle\hat{a}_{j}\rangle$ to $\boldsymbol{\phi}$. We do this by considering the Hellmann-Feynman theorem \cite{GriffithsBook} within the MF approximation, which states
\begin{equation}
\frac{\partial e_{\text{MF}}}{\partial\phi_{j}} =\left\langle\frac{\partial\hat{h}_{\text{MF}}}{\partial\phi_{j}}\right\rangle.
\end{equation}
Evaluating the right-hand side from (\ref{hMF}) and the left-hand side from (\ref{eMFapprox}), we obtain
\begin{subequations}\label{Hellmann-Feynman}
\begin{align}
\langle\hat{a}_{\text{o}}\rangle & =\frac{2J}{U}\chi_{\text{o}}\phi_{\text{e}},
\\
\langle\hat{a}_{\text{e}}\rangle & =\frac{2J}{U}\chi_{\text{e}}\phi_{\text{o}},
\end{align}
\end{subequations}
to leading order in $J$. Putting expressions (\ref{eMFapprox}) and (\ref{Hellmann-Feynman}) together in (\ref{e_gen}), and keeping terms only up to quadratic order in $\boldsymbol{\phi}$, we obtain the quadratic form
\begin{equation}
e \approx e_{\bar{n}_{\text{o}}\bar{n}_{\text{e}}}^{(0)}+\frac{4J^{2}}{U}[\chi_{\text{e}}|\phi_{\text{o}}|^{2}+\chi_{\text{o}}|\phi_{\text{e}}|^{2}-2J\chi_{\text{o}}\chi_{\text{e}}(\phi_{\text{e}}^{*}\phi_{\text{o}}+\phi_{\text{e}}\phi_{\text{o}}^{*})]=e_{\bar{n}_{\text{o}}\bar{n}_{\text{e}}}^{(0)}+\boldsymbol{\phi}^{\dagger}\mathcal{K}\boldsymbol{\phi},
\end{equation}
with real, symmetric curvature matrix
\begin{equation}
\mathcal{K}=\frac{4J^{2}}{U}\left(\begin{array}{cc}
\chi_{\text{e}} & -2J\chi_{\text{o}}\chi_{\text{e}}/U
\\
-2J\chi_{\text{o}}\chi_{\text{e}}/U & \chi_{\text{o}}
\end{array}\right).
\end{equation}
We can diagonalize the curvature matrix as $\mathcal{K}=\mathcal{S}^{\mathsf{T}}\text{diag}(\kappa_{+},\kappa_{-})\mathcal{S}$, where $\mathcal{S}$ is an orthonormal matrix (whose specific form is irrelevant for our purposes) and the eigenvalues read
\begin{equation}
\kappa_{\pm}=\frac{2J^{2}}{U}\left(\chi_{\text{e}}+\chi_{\text{o}}\pm\sqrt{(\chi_{\text{e}}-\chi_{\text{o}})^{2}+16J^{2}\chi_{\text{o}}^{2}\chi_{\text{e}}^{2}/U^{2}}\right).
\end{equation}
Defining then new complex parameters $(\phi_{+},\phi_{-})^{\mathsf{T}}=\mathcal{S}\boldsymbol{\phi}$, the `Landau potential' $e(\boldsymbol{\phi})$ can be written as
\begin{equation}
e(\boldsymbol{\phi}) \approx e_{0}+\kappa_{+}|\phi_{+}|^{2}+\kappa_{-}|\phi_{-}|^{2}.
\end{equation}
Therefore, the condition for stability of the $\boldsymbol{\phi}=0$ configuration is $\kappa_{\pm}>0$ (which $J=0$ satisfies), while as soon as the curvature $\kappa_{-}$ turns negative, that trivial configuration becomes unstable (which happens for sufficiently large $J$). Hence, the boundary between insulator and superfluid phases is determined by the condition
\begin{equation}
\kappa_{-}=0\;\;\;\Longleftrightarrow\;\;\;\frac{J}{U}=\frac{1}{2\sqrt{\chi_{\text{o}}\chi_{\text{e}}}}.\label{InsuSuperfluidBoundaryAnalytic}
\end{equation}
As shown in Fig. \ref{fig:MF_SM}, this simple analytic condition fits perfectly the insulating/superfluid boundaries found via self-consistent Hartree-Fock in the previous section.

\subsection{Perturbation theory on the MF nonlinear eigenproblem}

Before proceeding to the next section, we need to present the perturbative solution to the nonlinear eigenproblem defined by $\hat{h}_{\text{MF}}(\langle\hat{n}_{\text{o}}\rangle,\langle\hat{n}_{\text{e}}\rangle)$, proving that it leads to Eq. (\ref{eMFapprox}). We start by writing the Hamiltonian as $\hat{h}_{\text{MF}}=\hat{h}_{0}+\hat{w}$, where
\begin{equation}
\hat{h}_{0} = (-\mu+2V\langle\hat{n}_{\text{e}}\rangle_{J=0})\hat{n}_{\text{o}}+(-\mu+2V\langle\hat{n}_{\text{o}}\rangle_{J=0})\hat{n}_{\text{e}}+\frac{U}{2}[\hat{n}_{\text{o}}(\hat{n}_{\text{o}}-1)+\hat{n}_{\text{e}}(\hat{n}_{\text{e}}-1)]-2V\langle\hat{n}_{\text{o}}\rangle_{J=0}\langle\hat{n}_{\text{e}}\rangle_{J=0},
\end{equation}
is the MF Hamiltonian per pair of sites at $J=0$, which we take as our unperturbed Hamiltonian, and 
\begin{align}
\hat{w}= & -2J(\phi_{\text{e}}^{*}\hat{a}_{\text{o}}+\phi_{\text{e}}\hat{a}_{\text{o}}^{\dagger}+\phi_{\text{o}}^{*}\hat{a}_{\text{e}}+\phi_{\text{o}}\hat{a}_{\text{e}}^{\dagger}-\phi_{\text{o}}^{*}\phi_{\text{e}}-\phi_{\text{e}}^{*}\phi_{\text{o}})
\\
 & +2V[(\langle\hat{n}_{\text{e}}\rangle-\langle\hat{n}_{\text{e}}\rangle_{J=0})\hat{n}_{\text{o}}+(\langle\hat{n}_{\text{o}}\rangle-\langle\hat{n}_{\text{o}}\rangle_{J=0})\hat{n}_{\text{e}}-(\langle\hat{n}_{\text{o}}\rangle\langle\hat{n}_{\text{e}}\rangle-\langle\hat{n}_{\text{o}}\rangle_{J=0}\langle\hat{n}_{\text{e}}\rangle_{J=0})],\nonumber 
\end{align}
is the perturbation. Even though $\hat{h}_{0}$ defines also a nonlinear problem, its eigenstates are trivially found to be the Fock states $\{|n_{\text{o}},n_{\text{e}}\rangle\equiv|n_{\text{o}}\rangle\otimes|n_{\text{e}}\rangle\}_{n_{j}=0,1,2,...}$, as it depends solely on number operators. We thus have $\hat{h}_{0}|n_{\text{o}},n_{\text{e}}\rangle=e_{n_{\text{o}}n_{\text{e}}}^{(0)}|n_{\text{o}},n_{\text{e}}\rangle$, with
\begin{equation}
e_{n_{\text{o}}n_{\text{e}}}^{(0)}=-\mu(n_{\text{o}}+n_{\text{e}})+2Vn_{\text{o}}n_{\text{e}}+\frac{U}{2}[n_{\text{o}}(n_{\text{o}}-1)+n_{\text{e}}(n_{\text{e}}-1)].
\end{equation}
These coincide with the eigenenergies of the full Hamiltonian per pair of sites $\hat{H}/(L/2)$ for $J=0=\varepsilon$. Hence, $|\bar{n}_{\text{o}},\bar{n}_{\text{e}}\rangle$, with Fock numbers provided by (\ref{n_GS}), is the ground state of $\hat{h}_{0}$. We already found that this ground-state energy has a degeneracy related to the translational invariance of the model. In addition, for $2V>U$ it is also degenerate when $\mu/U$ is a positive integer or zero, leading to the same energies for $|\bar{n}_{\text{o}},0\rangle$ and $|\bar{n}_{\text{o}}+1,0\rangle$. Similarly, for $2V<U$ the degeneracy appears when $\mu/(U+2V)$ is a positive integer or zero. Nevertheless, we will see that non-degenerate perturbation theory gives sensible results even at those degeneracy points, so there is no need to treat them separately with degenerate perturbation theory.

We consider the full MF eigenproblem
\begin{equation}
\hat{h}_{\text{MF}}|\psi_{n_{\text{o}}n_{\text{e}}}\rangle=e_{n_{\text{o}}n_{\text{e}}}|\psi_{n_{\text{o}}n_{\text{e}}}\rangle,\label{MFeigenproblem}
\end{equation}
and assume as usual in perturbation theory that the eigenstates and eigenenergies admit a series expansion in $J$
\begin{subequations}\label{EigenPerturbativeExpansion}
\begin{align}
|\psi_{n_{\text{o}}n_{\text{e}}}\rangle & =|n_{\text{o}},n_{\text{e}}\rangle+J|\psi_{n_{\text{o}}n_{\text{e}}}^{(1)}\rangle+J^{2}|\psi_{n_{\text{o}}n_{\text{e}}}^{(2)}\rangle+...,
\\
e_{n_{\text{o}}n_{\text{e}}} & =e_{n_{\text{o}}n_{\text{e}}}^{(0)}+Je_{n_{\text{o}}n_{\text{e}}}^{(1)}+J^{2}e_{n_{\text{o}}n_{\text{e}}}^{(2)}+...\;\;\;.\label{EigenenergiesExpansion}
\end{align}
\end{subequations}
As we will see, we can choose $\langle n_{\text{o}},n_{\text{e}}|\psi_{n_{\text{o}}n_{\text{e}}}^{(1)}\rangle\in\mathbb{R}$. Moreover, we can demand that the eigenstates are normalized at all orders, so that
\begin{align}
1=\langle\psi_{n_{\text{o}}n_{\text{e}}}|\psi_{n_{\text{o}}n_{\text{e}}}\rangle= & \overbrace{\langle n_{\text{o}},n_{\text{e}}|n_{\text{o}},n_{\text{e}}\rangle}^{1}+J(\langle n_{\text{o}},n_{\text{e}}|\psi_{n_{\text{o}}n_{\text{e}}}^{(1)}\rangle+\langle\psi_{n_{\text{o}}n_{\text{e}}}^{(1)}|n_{\text{o}},n_{\text{e}}\rangle)
\\
& +J^{2}(\langle n_{\text{o}},n_{\text{e}}|\psi_{n_{\text{o}}n_{\text{e}}}^{(2)}\rangle+\langle\psi_{n_{\text{o}}n_{\text{e}}}^{(2)}|n_{\text{o}},n_{\text{e}}\rangle+\langle\psi_{n_{\text{o}}n_{\text{e}}}^{(1)}|\psi_{n_{\text{o}}n_{\text{e}}}^{(1)}\rangle)+...,\nonumber 
\end{align}
implies
\begin{subequations}
\begin{align}
\langle n_{\text{o}},n_{\text{e}}|\psi_{n_{\text{o}}n_{\text{e}}}^{(1)}\rangle & =0,
\\
\langle n_{\text{o}},n_{\text{e}}|\psi_{n_{\text{o}}n_{\text{e}}}^{(2)}\rangle+\langle\psi_{n_{\text{o}}n_{\text{e}}}^{(2)}|n_{\text{o}},n_{\text{e}}\rangle & =-\langle\psi_{n_{\text{o}}n_{\text{e}}}^{(1)}|\psi_{n_{\text{o}}n_{\text{e}}}^{(1)}\rangle.
\end{align}
\end{subequations}
These expressions will be useful later. Next we evaluate the expectation values appearing in the Hamiltonian:
\begin{align}
\langle\hat{n}_{j}\rangle= & \overbrace{\langle n_{\text{o}},n_{\text{e}}|\hat{n}_{j}|n_{\text{o}},n_{\text{e}}\rangle}^{n_{j}=\langle\hat{n}_{j}\rangle_{J=0}}+J(\overbrace{\langle n_{\text{o}},n_{\text{e}}|\hat{n}_{j}|\psi_{n_{\text{o}}n_{\text{e}}}^{(1)}\rangle}^{n_{j}\langle n_{\text{o}},n_{\text{e}}|\psi_{n_{\text{o}}n_{\text{e}}}^{(1)}\rangle=0}+\overbrace{\langle\psi_{n_{\text{o}}n_{\text{e}}}^{(1)}|\hat{n}_{j}|n_{\text{o}},n_{\text{e}}\rangle}^{n_{j}\langle\psi_{n_{\text{o}}n_{\text{e}}}^{(1)}|n_{\text{o}},n_{\text{e}}\rangle=0})
\\
& +J^{2}(\underbrace{\langle n_{\text{o}},n_{\text{e}}|\hat{n}_{j}|\psi_{n_{\text{o}}n_{\text{e}}}^{(2)}\rangle+\langle\psi_{n_{\text{o}}n_{\text{e}}}^{(2)}|\hat{n}_{j}|n_{\text{o}},n_{\text{e}}\rangle}_{-n_{j}\langle\psi_{n_{\text{o}}n_{\text{e}}}^{(1)}|\psi_{n_{\text{o}}n_{\text{e}}}^{(1)}\rangle}+\langle\psi_{n_{\text{o}}n_{\text{e}}}^{(1)}|\hat{n}_{j}|\psi_{n_{\text{o}}n_{\text{e}}}^{(1)}\rangle)+...,\nonumber 
\end{align}
so
\begin{subequations}
\begin{align}
\langle\hat{n}_{j}\rangle-\langle\hat{n}_{j}\rangle_{J=0} & =J^{2}\langle\psi_{n_{\text{o}}n_{\text{e}}}^{(1)}|(\hat{n}_{j}-n_{j})|\psi_{n_{\text{o}}n_{\text{e}}}^{(1)}\rangle+...
\\
\langle\hat{n}_{\text{o}}\rangle\langle\hat{n}_{\text{e}}\rangle-\langle\hat{n}_{\text{o}}\rangle_{J=0}\langle\hat{n}_{\text{e}}\rangle_{J=0} & =J^{2}[n_{\text{o}}\langle\psi_{n_{\text{o}}n_{\text{e}}}^{(1)}|(\hat{n}_{\text{e}}-n_{\text{e}})|\psi_{n_{\text{o}}n_{\text{e}}}^{(1)}\rangle+n_{\text{e}}\langle\psi_{n_{\text{o}}n_{\text{e}}}^{(1)}|(\hat{n}_{\text{o}}-n_{\text{o}})|\psi_{n_{\text{o}}n_{\text{e}}}^{(1)}\rangle]+...\;\;.
\end{align}
\end{subequations}
It is then convenient to split the perturbation as $\hat{w}=J\hat{h}+J^{2}\hat{v}$, with
\begin{subequations}
\begin{align}
\hat{h} & =-2(\phi_{\text{e}}^{*}\hat{a}_{\text{o}}+\phi_{\text{e}}\hat{a}_{\text{o}}^{\dagger}+\phi_{\text{o}}^{*}\hat{a}_{\text{e}}+\phi_{\text{o}}\hat{a}_{\text{e}}^{\dagger}-\phi_{\text{o}}^{*}\phi_{\text{e}}-\phi_{\text{e}}^{*}\phi_{\text{o}}),
\\
\hat{v} & =2V[\langle\psi_{n_{\text{o}}n_{\text{e}}}^{(1)}|(\hat{n}_{\text{e}}-n_{\text{e}})|\psi_{n_{\text{o}}n_{\text{e}}}^{(1)}\rangle(\hat{n}_{\text{o}}-n_{\text{o}})+\langle\psi_{n_{\text{o}}n_{\text{e}}}^{(1)}|(\hat{n}_{\text{o}}-n_{\text{o}})|\psi_{n_{\text{o}}n_{\text{e}}}^{(1)}\rangle(\hat{n}_{\text{e}}-n_{\text{e}})],
\end{align}
\end{subequations}
so that the MF Hamiltonian is written as
\begin{equation}
\hat{h}_{\text{MF}}=\hat{h}_{0}+J\hat{h}+J^{2}\hat{v},\label{hMF_Orders}
\end{equation}
with all terms showing explicitly the order in $J$ to which they contribute. Inserting (\ref{hMF_Orders}) and (\ref{EigenPerturbativeExpansion}) in (\ref{MFeigenproblem}), and matching like orders in $J$, we obtain\begin{subequations}
\begin{align}
\hat{h}_{0}|n_{\text{o}},n_{\text{e}}\rangle & =e_{n_{\text{o}}n_{\text{e}}}^{(0)}|n_{\text{o}},n_{\text{e}}\rangle,
\\
\hat{h}_{0}|\psi_{n_{\text{o}}n_{\text{e}}}^{(1)}\rangle+\hat{h}|n_{\text{o}},n_{\text{e}}\rangle & =e_{n_{\text{o}}n_{\text{e}}}^{(0)}|\psi_{n_{\text{o}}n_{\text{e}}}^{(1)}\rangle+e_{n_{\text{o}}n_{\text{e}}}^{(1)}|n_{\text{o}},n_{\text{e}}\rangle,\label{order1}
\\
\hat{h}_{0}|\psi_{n_{\text{o}}n_{\text{e}}}^{(2)}\rangle+\hat{h}|\psi_{n_{\text{o}}n_{\text{e}}}^{(1)}\rangle+\hat{v}|n_{\text{o}},n_{\text{e}}\rangle & =e_{n_{\text{o}}n_{\text{e}}}^{(0)}|\psi_{n_{\text{o}}n_{\text{e}}}^{(2)}\rangle+e_{n_{\text{o}}n_{\text{e}}}^{(1)}|\psi_{n_{\text{o}}n_{\text{e}}}^{(1)}\rangle+e_{n_{\text{o}}n_{\text{e}}}^{(2)}|n_{\text{o}},n_{\text{e}}\rangle,\label{order2}
\\
 & \vdots\nonumber 
\end{align}
\end{subequations}
The first identity is just the unperturbed eigenproblem. On the other hand, projecting (\ref{order1}) onto $|n_{\text{o}},n_{\text{e}}\rangle$, and using $\hat{a}_{j}|n_{j}\rangle=\sqrt{n_{j}}|n_{j}-1\rangle$ and $\hat{a}_{j}^{\dagger}|n_{j}\rangle=\sqrt{n_{j}+1}|n_{j}+1\rangle$, we obtain the first-order correction to the eigenenergies,
\begin{equation}
e_{n_{\text{o}}n_{\text{e}}}^{(1)}=\langle n_{\text{o}},n_{\text{e}}|\hat{h}|n_{\text{o}},n_{\text{e}}\rangle=2(\phi_{\text{o}}^{*}\phi_{\text{e}}-\phi_{\text{e}}^{*}\phi_{\text{o}}),
\end{equation}
while projecting (\ref{order1}) onto $|m_{\text{o}}\neq n_{\text{o}},m_{\text{e}}\neq n_{\text{e}}\rangle$, we find the first-order corrections to the eigenvectors,
\begin{equation}
\langle m_{\text{o}},m_{\text{e}}|\psi_{n_{\text{o}}n_{\text{e}}}^{(1)}\rangle=\frac{h_{m_{\text{o}}m_{\text{e}},n_{\text{o}}n_{\text{e}}}}{e_{n_{\text{o}}n_{\text{e}}}^{(0)}-e_{m_{\text{o}}m_{\text{e}}}^{(0)}},
\end{equation}
with
\begin{align}
h_{m_{\text{o}}m_{\text{e}},n_{\text{o}}n_{\text{e}}}=\langle m_{\text{o}},m_{\text{e}}|\hat{h}|n_{\text{o}},n_{\text{e}}\rangle=-2 & (\sqrt{n_{\text{o}}}\phi_{\text{e}}^{*}\delta_{m_{\text{o}},n_{\text{o}}-1}\delta_{m_{\text{e}},n_{\text{e}}}+\sqrt{n_{\text{o}}+1}\phi_{\text{e}}\delta_{m_{\text{o}},n_{\text{o}}+1}\delta_{m_{\text{e}},n_{\text{e}}}
\\
& +\sqrt{n_{\text{e}}}\phi_{\text{o}}^{*}\delta_{m_{\text{o}},n_{\text{o}}}\delta_{m_{\text{e}},n_{\text{e}}-1}+\sqrt{n_{\text{e}}+1}\phi_{\text{o}}\delta_{m_{\text{o}},n_{\text{o}}}\delta_{m_{\text{e}},n_{\text{e}}+1}).\nonumber 
\end{align}
Finally, the second-order correction to the eigenenergies is found by projecting (\ref{order2}) onto $|n_{\text{o}},n_{\text{e}}\rangle$, leading to
\begin{equation}
e_{n_{\text{o}}n_{\text{e}}}^{(2)}=\langle n_{\text{o}},n_{\text{e}}|\hat{h}|\psi_{n_{\text{o}}n_{\text{e}}}^{(1)}\rangle+\langle n_{\text{o}},n_{\text{e}}|\hat{v}|n_{\text{o}},n_{\text{e}}\rangle=\sum_{m_{\text{o}}\neq n_{\text{o}}}\sum_{m_{\text{e}}\neq n_{\text{e}}}\frac{|h_{m_{\text{o}}m_{\text{e}},n_{\text{o}}n_{\text{e}}}|^{2}}{e_{n_{\text{o}}n_{\text{e}}}^{(0)}-e_{m_{\text{o}}m_{\text{e}}}^{(0)}},
\end{equation}
where we have used
\begin{subequations}
\begin{align}
\langle n_{\text{o}},n_{\text{e}}|\hat{h}|\psi_{n_{\text{o}}n_{\text{e}}}^{(1)}\rangle & =\sum_{m_{\text{o}}\neq n_{\text{o}}}\sum_{m_{\text{e}}\neq n_{\text{e}}}\frac{h_{m_{\text{o}}m_{\text{e}},n_{\text{o}}n_{\text{e}}}}{e_{n_{\text{o}}n_{\text{e}}}^{(0)}-e_{m_{\text{o}}m_{\text{e}}}^{(0)}}\underbrace{\langle n_{\text{o}},n_{\text{e}}|\hat{h}|m_{\text{o}},m_{\text{e}}\rangle}_{h_{m_{\text{o}}m_{\text{e}},n_{\text{o}}n_{\text{e}}}^{*}},\\
\langle n_{\text{o}},n_{\text{e}}|\hat{v}|n_{\text{o}},n_{\text{e}}\rangle & \sim\langle n_{\text{o}},n_{\text{e}}|(\hat{n}_{j}-n_{j})|n_{\text{o}},n_{\text{e}}\rangle=0.
\end{align}
\end{subequations}
Recombining the eigenenergies up to order $J^{2}$ as in (\ref{EigenenergiesExpansion}) and particularizing to the ground state ($n_{\text{o}}=\bar{n}_{\text{o}}$,$n_{\text{e}}=\bar{n}_{\text{e}}$), we obtain the perturbative expression (\ref{eMFapprox}) that we wanted to prove.

\section{Coherent State Ansatz}\label{sec:Appendix-Coherent-State}

\subsection{Analytic solutions and superfluid/supersolid phase boundary}

As explained in the main text, deep in the region with $Z_{2}$-superfluid order we expect the amplitudes $\langle\hat{a}_{j}\rangle\neq0$ to be dominated by a strong coherent component. Hence, in this region it makes sense to consider a coherent-state ansatz \cite{CNB-QOnotes} of the form $\bigotimes_{j=1}^{L}|\alpha_{j}\rangle$, where $\hat{a}_{j}|\alpha_{j}\rangle=\alpha_{j}|\alpha_{j}\rangle$ and $\{\alpha_{j}\in\mathbb{C}\}_{j=1,2,...,L}$ are the variational parameters. Considering the Hamiltonian (\ref{H_SM}), the energy functional $\langle\hat{H}\rangle$ takes the form
\begin{equation}
E =\sum_{j=1}^{L}\left[-\mu\vert\alpha_{j}\vert^{2}-J(\alpha_{j}\alpha_{j+1}^{*}+\alpha_{j}^{*}\alpha_{j+1})-\frac{\varepsilon}{2}(\alpha_{j}^{2}+\alpha_{j}^{*2})+\frac{U}{2}\vert\alpha_{j}\vert^{4}+V\vert\alpha_{j}\vert^{2}\vert\alpha_{j+1}\vert^{2}\right].\label{CoherentEnergyFunctional}
\end{equation}
The $J$ term is minimized when all sites have the same phase, so that $\text{Re}\{\alpha_{j}\alpha_{j+1}^{*}\}\geq0$. On the other hand, in order to minimize the $\varepsilon$ term, the amplitudes $\alpha_{j}$ must be real. Hence, we conclude that the amplitudes are real and either all positive or all negative. On the other hand, since the energy functional is invariant under the $Z_{2}$ transformation $\alpha_{j}\rightarrow-\alpha_{j}$ $\forall j$, we can take the positive sign for the amplitudes for definiteness, just keeping in mind that from any minimizing configuration $\{\alpha_{j}\geq0\}_{j=1,2,...,L}$ we can build an equally valid configuration simply inverting the sign of all amplitudes. The minimization condition becomes then
\begin{equation}
\frac{\partial E}{\partial\alpha_{j}}=0\;\;\;\Rightarrow\;\;\;[-(\mu+\varepsilon)+U\alpha_{j}^{2}+V\alpha_{j+1}^{2}+V\alpha_{j-1}^{2}]\alpha_{j}=J(\alpha_{j+1}+\alpha_{j-1}),\label{CoherentConfigurationsCondition}
\end{equation}
with $\alpha_{j}$ real and positive $\forall j$. These equations possess three types of analytic solutions, which happen to be the only relevant minima according to our exhaustive exploration, as we discuss in the next section. The simplest among these is the trivial one, $\alpha_{j}=0$ $\forall j$, with null energy $E=0$ in all parameter space. One can also easily find in (\ref{CoherentConfigurationsCondition}) a nontrivial homogeneous solution (that we simply denote by `$Z_{2}$-superfluid' or SF in this coherent context)
\begin{equation}
\text{\ensuremath{\alpha_{j}=\sqrt{\frac{\mu+\varepsilon+2J}{U+2V}}}\ensuremath{\equiv\alpha_{\text{SF}}}}\;\forall j,
\end{equation}
with energy
\begin{equation}
E=-2LJ^{2}\frac{(\nu+1)^{2}}{U+2V}\equiv E_{\text{SF}}<0,\label{ESF}
\end{equation}
where we have defined the parameter $\nu=(\mu+\varepsilon)/2J$ that will reappear throughout this section on occasion. This SF solution has then lower energy than the trivial one. If we would allow for negative chemical potentials, the superfluid solution would cease to exist as soon as $\mu<-(\varepsilon+2J)$ or $\nu<-1$, parameter region where the trivial solution takes over (actually, for $\nu<-1$ the trivial solution is a minimum, while for $\nu>-1$ it becomes an unstable saddle point). However, this is not relevant to our current work where we are analyzing the $\mu>0$ region, since this is the usual accessible region in condensed matter physics.

The third analytic solution is the staggered one (that we denote by `$Z_{2}$-supersolid' or SS)
\begin{equation}
\alpha_{j}=\left\{ \begin{array}{cl}
\alpha_{\text{SS}} & \text{for }j\in\text{odd}
\\
r\alpha_{\text{SS}} & \text{for }j\in\text{even}
\end{array}\right.,
\end{equation}
where we assume that the density is larger at odd sites ($0\leq r\leq1$) for definiteness, but the opposite case is also a solution, owed to the translational invariance of the Hamiltonian. Particularizing (\ref{CoherentConfigurationsCondition}) to odd $j$ leads to
\begin{align}
\alpha_{\text{SS}}^{2} & =\frac{\mu+\varepsilon+2rJ}{U+2r^{2}V},
\end{align}
which inserted into (\ref{CoherentConfigurationsCondition}) particularized now to even $j$ leads to an equation for $r$ that reads
\begin{equation}
(1-r^{2})(1+r^{2}-ar)=0,
\end{equation}
where we have defined $a=\nu(2V/U-1)$. Apart from the $r=1$ root that simply leads to the SF solution discussed above, we obtain only another root satisfying $0\leq r\leq1$, namely $2r=a-\sqrt{a^{2}-4}$ that exists only for $a\geq2$, condition that can be recasted as
\begin{equation}
\frac{V}{U}\geq\frac{1}{2}+\frac{1}{\nu}\;\;\;\Leftrightarrow\;\;\;\mu\geq\frac{4J}{2V/U-1}-\varepsilon.
\end{equation}
Let us remark that the equality in this expression leads precisely to the SF/SS boundary that we have presented in the main text. Indeed, for $a=2$ we get $r=1$, meaning that the staggered solution converges to the homogeneous one at the boundary. Moreover, inserting the staggered solution into the coherent energy functional (\ref{CoherentEnergyFunctional}) leads to
\begin{equation}
E=-\frac{LJ^{2}}{U}\left(\nu^{2}+\frac{2}{2V/U-1}\right)\equiv E_{\text{SS}},\label{ESS}
\end{equation}
which is easily shown to be smaller than $E_{\text{SF}}$ in all the domain of existence of the solution. Hence, we conclude that $a=2$ is the coherent-state prediction for the phase boundary between the SF and SS regions. This method predicts a continuous second-order phase transition, since $E_{\text{SF}}$ and $E_{\text{SS}}$ and their first-order derivatives are continuously connected at the phase boundaries, with the discontinuity appearing at the second-order derivative.

\begin{figure}[t]
\includegraphics[width=0.6\columnwidth]{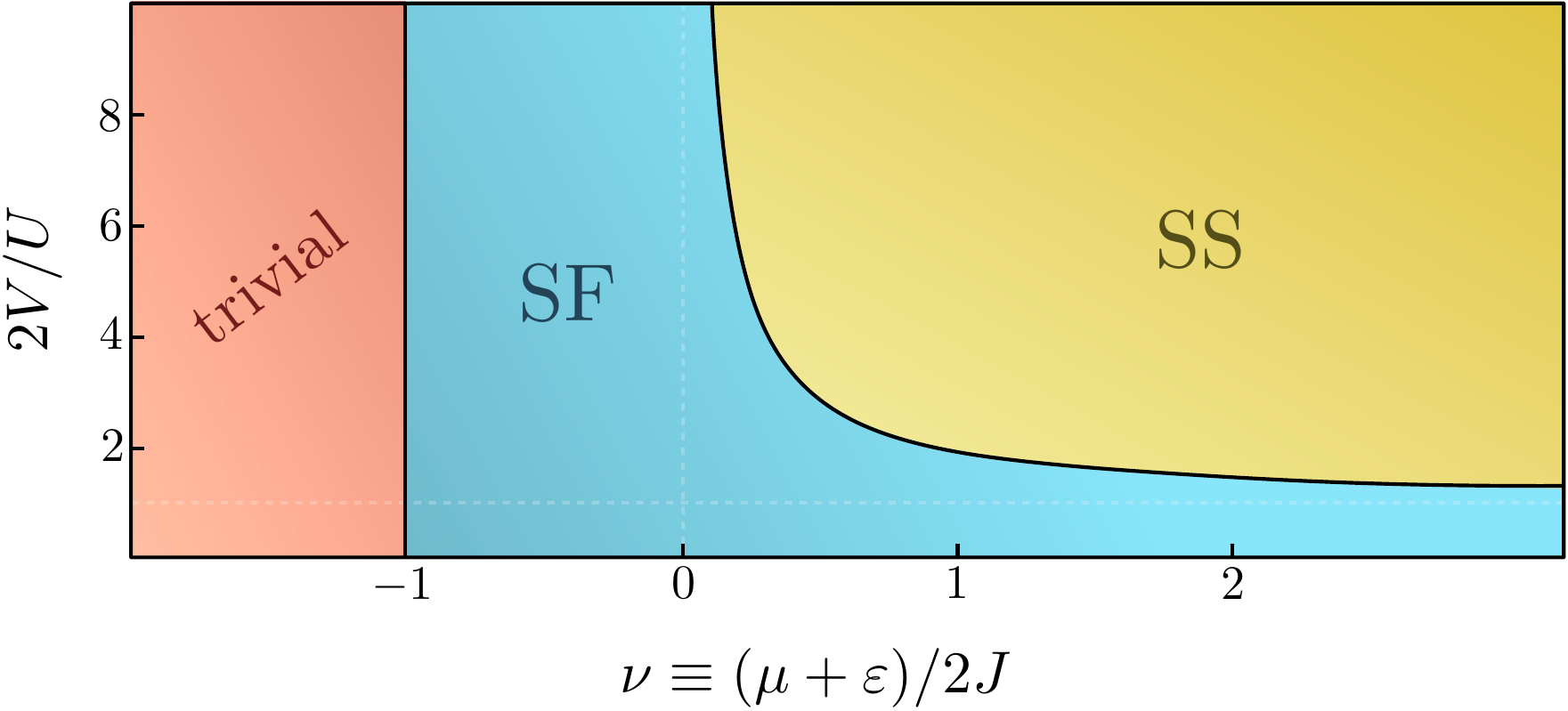} \caption{Phase diagram arising from the coherent-state ansatz, which is fully characterized by two reduced parameters only, $\nu=(\mu+\varepsilon)/2J$ and $V/U$. The trivial phase refers the $\alpha_{j}=0$ configuration for all coherent amplitudes (vacuum state). SF refers to a configuration with all amplitudes are equal and non-zero (homogeneous superfluid-like). SS refers to a configuration with alternating amplitudes, different for even and odd sites (supersolid-like). The boundary between the trivial and SF phases is $\nu=-1$. The boundary between SF and SS phases is $2V/U=1+2/\nu$, and exists only for $\nu>0$ and $2V/U>1$.} \label{fig:Coherent}
\end{figure}

The coherent-state ansatz provides then a phase diagram that is easily summarized in terms of just two relevant parameters $\nu$ and $2V/U$, as we show in Fig. \ref{fig:Coherent}.

\subsection{Numerical search of other minima\label{Sec:iTimeCoherent}}

The question remains about whether there exist other solutions to (\ref{CoherentConfigurationsCondition}) with an even lower energy. We have performed an exhaustive numerical search, finding many local minima, but never with energies below $E_{\text{SF}}$ or $E_{\text{SS}}$ of Eqs. (\ref{ESF}) and (\ref{ESS}), respectively. In this section we want to present the way in which we have performed the numerical
search.

We have relied on imaginary-time evolution. In particular, we know that the energy of the state
\begin{equation}
|\psi(\tau)\rangle=\frac{e^{-\hat{H}\tau}|\psi(0)\rangle}{\langle\psi(0)|e^{-2\hat{H}\tau}|\psi(0)\rangle},\label{ImTimeEvo}
\end{equation}
should converge at $\tau\rightarrow\infty$ to that of the system's ground state, assuming that $|\psi(0)\rangle$ projects onto the ground subspace. The differential form of (\ref{ImTimeEvo}) is
\begin{equation}
\partial_{\tau}|\psi(\tau)\rangle=-\left[\hat{H}-\langle\psi(\tau)|\hat{H}|\psi(\tau)\rangle\right]|\psi(\tau)\rangle,\label{ImTimeEq}
\end{equation}
which has the form of a (non-linear) Schr\"odinger equation where time has been replaced by an imaginary time $-\mathrm{i}\tau$. We can particularize this equation to the coherent-state manifold by considering the ansatz of the previous section,
\begin{equation}
|\psi(\tau)\rangle=\bigotimes_{j=1}^{L}|\alpha_{j}(\tau)\rangle,\label{CoherentStates_tau}
\end{equation}
but now with $\tau$-dependent coherent amplitudes $\alpha_{j}(\tau)$. We can then turn (\ref{ImTimeEq}) into an evolution or `flow' equation for these amplitudes by considering
\begin{align}
\partial_{\tau}\langle\psi(\tau)|\hat{a}_{j}|\psi(\tau)\rangle & =\left[\partial_{\tau}\langle\psi(\tau)|\right]\hat{a}_{j}|\psi(\tau)\rangle+\langle\psi(\tau)|\hat{a}_{j}\left[\partial_{\tau}|\psi(\tau)\rangle\right]
\\
& =-\langle\psi(\tau)|\hat{H}\hat{a}_{j}+\hat{a}_{j}\hat{H}|\psi(\tau)\rangle+2\langle\psi(\tau)|\hat{H}|\psi(\tau)\rangle\langle\psi(\tau)|\hat{a}_{j}|\psi(\tau)\rangle\nonumber
\\
& =-\langle\psi(\tau)|[\hat{a}_{j},\hat{H}]|\psi(\tau)\rangle+2\Big[\langle\psi(\tau)|\hat{H}\underbrace{\hat{a}_{j}|\psi(\tau)\rangle}_{\alpha_{j}|\psi(\tau)\rangle}-\langle\psi(\tau)|\hat{H}|\psi(\tau)\rangle\underbrace{\langle\psi(\tau)|\hat{a}_{j}|\psi(\tau)\rangle}_{\alpha_{j}}\Big],\nonumber 
\end{align}
so that (here, the dot denotes derivative with respecto to $\tau$)
\begin{equation}
\dot{\alpha}_{j}=-\langle[\hat{a}_{j},\hat{H}]\rangle=\left[\mu-U\vert\alpha_{j}\vert^{2}+V\left(\vert\alpha_{j+1}\vert^{2}+\vert\alpha_{j-1}\vert^{2}\right)\right]\alpha_{j}+\varepsilon\alpha_{j}^{*}+J(\alpha_{j+1}+\alpha_{j-1}),
\end{equation}
where, after performing the commutator, we have used
\begin{equation}
\langle\hat{a}_{j_{1}}^{\dagger}\hat{a}_{j_{2}}^{\dagger}...\hat{a}_{j_{J}}^{\dagger}\hat{a}_{j_{J+1}}\hat{a}_{j_{J+2}}...\hat{a}_{j_{J+K}}\rangle=\alpha_{j_{1}}^{*}\alpha_{j_{2}}^{*}...\alpha_{j_{J}}^{*}\alpha_{j_{J+1}}\alpha_{j_{J+2}}...\alpha_{j_{J+K}},
\end{equation}
for the normally-ordered product of $J$ creation operators with $K$ annihilation operators in a coherent state. From any initial configuration $\{\alpha_{j}(0)\}_{j=1,2,...,L}$ (different than a fixed point), this nonlinear system will deterministically flow towards one of its attractors, corresponding to a local (or global) minimum of the energy functional of Eq. (\ref{CoherentEnergyFunctional}). We have tried a huge number of random initial configurations for sizes $L$ as large as few hundreds, always flowing either into the analytical solutions presented in the previous sections, or other configurations with higher energy.

\section{General Gaussian-state ansatz}\label{sec:Gaussian}

In order to confirm the predictions from the coherent-state ansatz, we now extend the variational space by considering all possible Gaussian states. We will show that the predictions from this general ansatz converge to the analytical ones obtained in the previous section as we move deeper into the region with $Z_{2}$-superfluid order.

Gaussian states refer to those with a Gaussian Wigner function \citep{CNB-QOnotes,CNBbook,Gaussian1,Gaussian2,ParisBook}, such that their statistics are completely characterized by first and second order moments. It is customary to refer these moments to the position and momentum quadratures $\hat{x}_{j}=\hat{a}_{j}^{\dagger}+\hat{a}_{j}$ and $\hat{p}_{j}=\mathrm{i}(\hat{a}_{j}^{\dagger}-\hat{a}_{j})$, rather than the annihilation and creation operators, which we collect in the column vector $\hat{\boldsymbol{r}}=(\hat{x}_{1},...,\hat{x}_{L},\hat{p}_{1},...,\hat{p}_{L})^{\mathsf{T}}$, where $\mathsf{T}$ transposes the array without altering the operators that form it. Note that the canonical commutation relations read
\begin{equation}
[\hat{r}_{m},\hat{r}_{n}]=2\mathrm{i}\Omega_{mn},\;\;\;\text{with }\Omega=\left(\begin{array}{cc}
0 & \mathcal{I}\\
-\mathcal{I} & 0
\end{array}\right),\label{CCR}
\end{equation}
with $\mathcal{I}$ the $L\times L$ identity matrix and $\Omega$ known as the symplectic form. First and second order moments are characterized, respectively, by the mean vector $d_{m}=\langle\hat{r}_{m}\rangle$ and the covariance matrix $V_{mn}=\langle\delta\hat{r}_{m}\delta\hat{r}_{n}+\delta\hat{r}_{n}\delta\hat{r}_{m}\rangle/2$, where we have defined the quadrature fluctuations $\delta\hat{r}_{m}=\hat{r}_{m}-d_{m}$. In a more compact notation
\begin{subequations}
\begin{align}
\boldsymbol{d} & =\langle\boldsymbol{\hat{r}}\rangle\in\mathbb{R}^{2L},
\\
V & =\langle\delta\boldsymbol{\hat{r}}\delta\boldsymbol{\hat{r}}^\mathsf{T}\rangle-\mathrm{i}\Omega\in\mathbb{R}^{2L}\times\mathbb{R}^{2L},\label{Vdef}
\end{align}
\end{subequations}
where we have used the commutation relations (\ref{fig:Coherent}) in the second line. Higher-order moments can be found from first and second order ones via the Gaussian moment theorem \cite{MandelWolfBook,CNB-QOnotes}. Consider a set of $K$ operators $\{\hat{L}_{k}\}_{k=1,2,...,K}$ all arbitrary linear functions of the quadrature fluctuations; this theorem states that
\begin{equation}
\langle\hat{L}_{1}\hat{L}_{2}...\hat{L}_{K}\rangle=\left\{ \begin{array}{cc}
0 & \text{for odd }K
\\
\sum_{\text{all }(K-1)!!\text{ pairings}}^{\{k_{1},k_{2},...,k_{K}\}\in}\langle\hat{L}_{k_{1}}\hat{L}_{k_{2}}\rangle...\langle\hat{L}_{k_{K-1}}\hat{L}_{k_{K}}\rangle & \text{for even }K
\end{array}\right..\label{GaussianMomentTheoremQuantum}
\end{equation}
Together with the relation $\langle\delta\hat{r}_{m}\delta\hat{r}_{n}\rangle=V_{mn}+\mathrm{i}\Omega_{mn}$, this allows connecting the expectation value of any polynomial function of the quadratures to a polynomial of the elements of the mean vector and covariance matrix of the Gaussian state. Crucially, since the Hamiltonian (\ref{H_SM}) is a fourth-order polynomial in the annihilation and creation operators (or the quadratures, since $2\hat{a}_{j}=\hat{x}_{j}+\mathrm{i}\hat{p}_{j}$), the energy functional $E(\boldsymbol{d},V)=\langle\hat{H}\rangle$ can be found as a function of the mean vector and covariance matrix by using this Gaussian moment theorem (\ref{GaussianMomentTheoremQuantum}). While this can be a tedious task, it is conceptually simple. The Binder parameter as defined in the main text is determined from the ratio between eighth-order polynomials in the quadratures, and it is another quantity that we will use extensively in this section and we compute via the Gaussian moment theorem.

It is not difficult to prove \citep{CNB-QOnotes,CNBbook,Gaussian1,Gaussian2,ParisBook} that any pure Gaussian state $|\psi(V,\boldsymbol{d})\rangle$ characterized by its mean vector and covariance matrix can be generated by applying some Gaussian unitary $\hat{G}$ onto the vacuum state
\begin{equation}
|\psi(V,\boldsymbol{d})\rangle=\hat{G}|vac\rangle,\;\;\;\text{with }\hat{a}_{j}|vac\rangle=0,\label{psi_Gaussian}
\end{equation}
where Gaussian unitaries are defined as those that perform a linear transformation on the quadratures
\begin{equation}
\hat{G}^{\dagger}\boldsymbol{\hat{r}}\hat{G}=S\boldsymbol{\hat{r}}+\boldsymbol{d}.\label{GaussianUnitary}
\end{equation}
Preservation of the canonical commutation relations imposes that the matrix $S$ is constrained to the special symplectic group of $L$ modes, $S_{p}(2L,\mathbb{R})$, that is, it has unit determinant and must leave invariant the symplectic form $\Omega$,
\begin{equation}
S\Omega S^{\mathsf{T}}=\Omega.
\end{equation}
Taking into account that the covariance matrix of vacuum is the $2L\times2L$ identity, (\ref{Vdef}), (\ref{psi_Gaussian}), and (\ref{GaussianUnitary}) imply the relation
\begin{equation}
V=SS^{\mathsf{T}},\label{StoV}
\end{equation}
between the covariance matrix of the Gaussian state and the Gaussian unitary that defines it. The coherent states of the previous section, Eq. (\ref{CoherentStates_tau}), correspond to Gaussian states with $V$ equal to the identity, $d_{j}=2\text{Re}\{\alpha_{j}\}$, and $d_{L+j}=2\text{Im}\{\alpha_{j}\}$.

In order to find the ground state within the Gaussian variational manifold we make use again of imaginary-time evolution. Considering a Gaussian state with $\tau$-dependent variational parameters $\boldsymbol{d}(\tau)$ and $V(\tau)$, it is possible to show \cite{Shi17} using (\ref{psi_Gaussian}) that the imaginary-time Schr\"odinger equation (\ref{ImTimeEq}) is transformed into
\begin{subequations}\label{FlowEqs}
\begin{align}
\dot{\boldsymbol{d}} & =-2V\frac{\delta E}{\delta\boldsymbol{d}},
\\
\dot{V} & =4\Omega^{T}\frac{\delta E}{\delta V}\Omega-4V\frac{\delta E}{\delta V}V,
\end{align}
\end{subequations}
where we have defined a vector and a matrix containing the derivatives of the energy functional with respect the variational parameters
\begin{subequations}
\begin{align}
\left(\frac{\delta E}{\delta\boldsymbol{d}}\right)_{m} & =\frac{\partial E}{\partial d_{m}},
\\
\left(\frac{\delta E}{\delta V}\right)_{mn} & =\frac{\partial E}{\partial V_{mn}}.
\end{align}
\end{subequations}
For a given parameter set, we simulate this `flow' equations (\ref{FlowEqs}) starting from many random mean vectors and covariance matrices, up to a sufficiently long time $\tau$ that ensures convergence to a fixed point of the equations (corresponding, as in the coherent case, to a local minimum of the energy landscape). We then select the one providing the lowest energy.

\begin{figure}[t]
\includegraphics[width=1\textwidth]{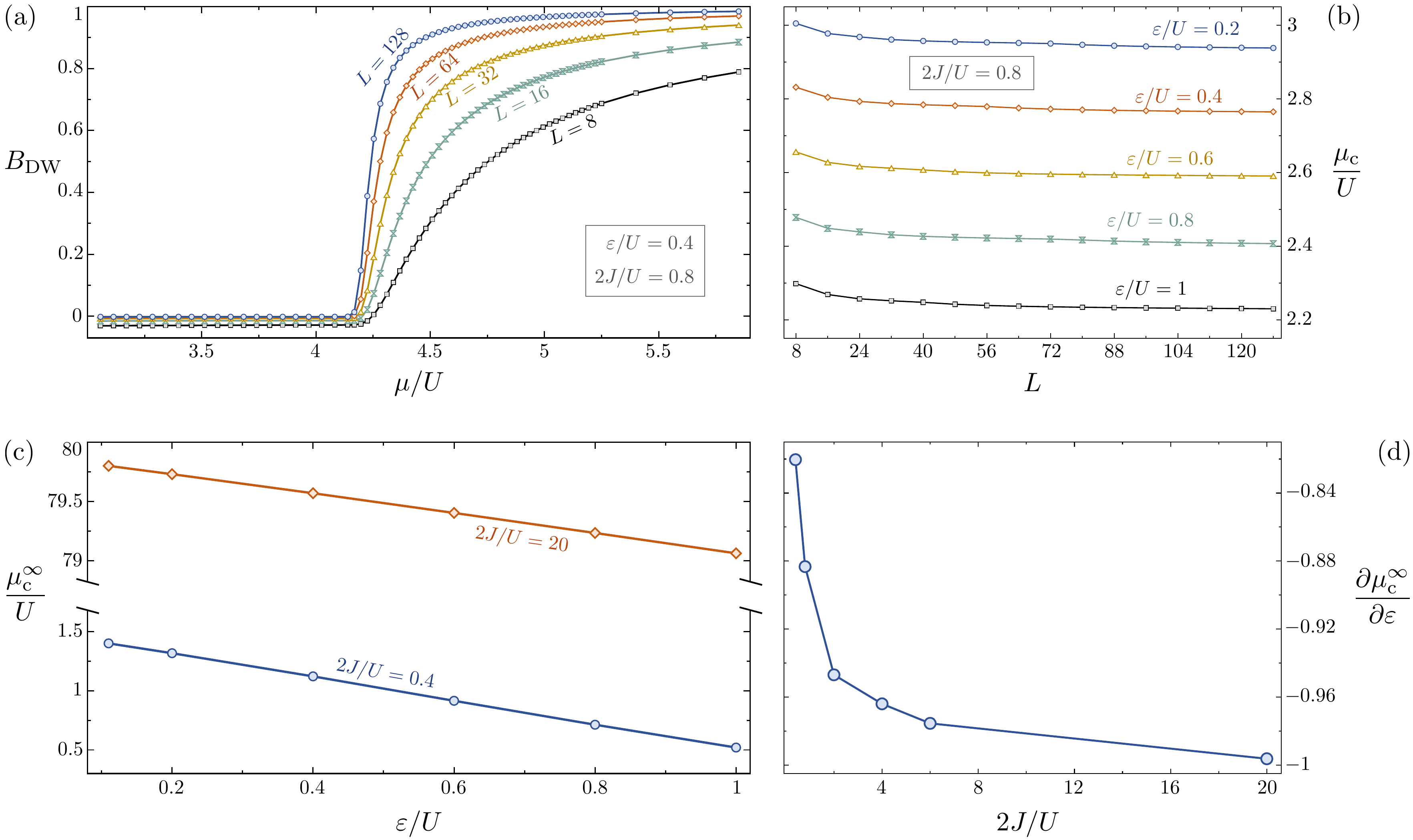} \caption{Results from the general Gaussian ground-state ansatz for $2V/U=1.5$. (a) Binder parameter $B$ as a function of $\mu/U$ for increasing system sizes $L$; $2J/U=0.8$ and $\varepsilon/U=0.4$ have been chosen for this
plot, but similar plots are emerge for other values. It is evident that a transition from homogeneous to staggered superfluid ($Z_{2}$-SF to $Z_{2}$-SS) is predicted at some critical chemical potential. (b) Largest value of the chemical potential for which the Binder parameter is zero, which we define as $\mu_{\text{c}}/U$, as a function of the system size $L$, for different values of $\varepsilon/U$ and $2J/U=0.8$. It is clear that $\mu_{\text{c}}/U$ converges towards some value $\mu_{\text{c}}^{\infty}/U$ as $L$ increases. In (c) we plot this value as a function of $\varepsilon/U$ for two values of $2J/U$. We can appreciate a linear relationship that is also found for any other value of $2J/U$. Finally, in (d) we plot the slope of the previous straight lines, $\partial\mu_{\text{c}}^{\infty}/\partial\varepsilon$, for different values of $2J/U$. The slope tends to $-1$ as $2J/U$ increases, confirming the coherent-state prediction $\mu_{\text{c}}\sim-\varepsilon$.} \label{fig:Gaussian}
\end{figure}

Before presenting the results, let us comment on one subtle point related to how we generate random (but physical) covariance matrices. We exploit the Bloch-Messiah reduction or Euler decomposition of symplectic matrices \citep{CNBbook,Gaussian1,ParisBook}, so that Eq. (\ref{StoV}) allows us to sample covariance matrices as
\begin{equation}
V=K^{\text{T}}DK,
\end{equation}
where $D=\text{diag(\ensuremath{e^{-2r_{1}}},...,\ensuremath{e^{-2r_{L}}},\ensuremath{e^{2r_{1}}},...,\ensuremath{e^{2r_{L}}})}$ is a diagonal matrix with real and positive random numbers $r_{j}$ (technically $D$ is the covariance matrix of $L$ independent squeezed vacuum states \citep{CNB-QOnotes,CNBbook,Gaussian1,Gaussian2,ParisBook}), and $K$ is a symplectic orthogonal matrix (so-called interferometer) that can be built as \citep{ParisBook}
\begin{equation}
K=\left(\begin{array}{cc}
X & -Y
\\
Y & X
\end{array}\right),
\end{equation}
being $X$ and $Y$ the real and imaginary parts of a random $L\times L$ unitary matrix $Q=X+\mathrm{i}Y$ that we can generate from the QR decomposition of any random complex matrix.

Our results are summarized in Fig. \ref{fig:Gaussian}. Our goal is showing that the general Gaussian ansatz is compatible with the coherent-state ansatz deep in the superfluid region. In particular, we aim at showing that there is an asymptotic $\mu\sim-\varepsilon$ relation between the chemical potential and the pair injection at the critical boundary separating the SF and SS phases as $2J/U$ increases. To this aim, we proceed as follows. We fix $2V/U=1.5$. Then, for any value of $2J/U$ and $\varepsilon/U$ we expect a transition from SF to SS for some value of $\mu/U$. We use the Binder parameter to perform a finite-size scaling and determine the critical $\mu/U$ as a function of $\varepsilon/U$ for different values of $J/U$. An example of the behavior of this parameter is shown in Fig. \ref{fig:Gaussian}a for $J/U=0.4=\varepsilon/U$ (similar figures are found for any other values). We check system sizes ranging from $L=8$ to 128, as specified in the figure. For all of them, the Binder parameter changes from zero (or right below zero), to a positive increasing function closer to a step function the larger $L$ is. For a given system size $L$, we then define the critical chemical potential $\mu_{\text{c}}$ as the largest $\mu$ for which $B$ is zero (we estimate it by interpolating the points via the Makima algorithm, consisting of piecewise third-order polynomials with continuous first-order derivatives). In Fig. \ref{fig:Gaussian}b we show this $\mu_{\text{c}}$ as a function of $L$ (for different values of $\varepsilon$ and $J=0.4$), which is expected to adhere to a power law of the type $\mu_{\text{c}}=\mu_{\text{c}}^{\infty}+\beta L^{-\eta}$ for some positive parameters $\{\mu_{\text{c}}^{\infty},\beta,\eta\}$. Indeed, we find that such curve fits nicely to our data, allowing us to estimate the critical chemical potential $\mu_{\text{c}}^{\infty}/U$
and find a nice linear dependence on $\varepsilon/U$, Fig. \ref{fig:Gaussian}c, for each value of $J/U$. Finally, we consider the slope $\partial\mu_{\text{c}}^{\infty}/\partial\varepsilon$ of those linear curves as a function of $J/U$, finding that it approaches $-1$ as $J/U$ increases, that is, as we move deeper into the superfluid region. This serves as support for the coherent-state predictions. 

\section{Model with full $U(1)$ symmetry}\label{sec:ModelU1}

The pair-injection term $-\sum_{j}\varepsilon(\hat{a}_{j}^{2}+\hat{a}_{j}^{\dagger2})/2$ considered in our work explicitly breaks the $U(1)$ symmetry of the extended Bose-Hubbard model, leaving only a $Z_{2}$ symmetry $\hat{a}_{j}\rightarrow-\hat{a}_{j}$ $\forall j$. As we mentioned in the conclusions, one can consider more elaborate extensions of the Bose-Hubbard model with pair-injection terms that still possess full $U(1)$ symmetry. Here we put forward one of such models and show that at the coherent-state-ansatz level, pair injection has the same effect as in the simpler model studied in the main text.

Consider a chain with two bosonic modes per site, with annihilation operators $\{\hat{a}_{j},\hat{b}_{j}\}_{j=1,2,...,L}$, satisfying canonical commutation relations, $[\hat{a}_{j},\hat{a}_{l}^{\dagger}]=\delta_{jl}=[\hat{b}_{j},\hat{b}_{l}^{\dagger}]$ with any other commutator vanishing. Each set of bosonic modes is subject to an extended Bose-Hubbard Hamiltonian,
\begin{subequations}
\begin{align}
\hat{H}_{a} & =\sum_{j=1}^{L}\left[-\mu\hat{a}_{j}^{\dagger}\hat{a}_{j}+\frac{U}{2}\hat{a}_{j}^{\dagger2}\hat{a}_{j}^{2}-J\left(\hat{a}_{j}^{\dagger}\hat{a}_{j+1}+\hat{a}_{j+1}^{\dagger}\hat{a}_{j}\right)+V\hat{a}_{j+1}^{\dagger}\hat{a}_{j+1}\hat{a}_{j}^{\dagger}\hat{a}_{j}\right],
\\
\hat{H}_{b} & =\sum_{j=1}^{L}\left[-\mu\hat{b}_{j}^{\dagger}\hat{b}_{j}+\frac{U}{2}\hat{b}_{j}^{\dagger2}\hat{b}_{j}^{2}-J\left(\hat{b}_{j}^{\dagger}\hat{b}_{j+1}+\hat{b}_{j+1}^{\dagger}\hat{b}_{j}\right)+V\hat{b}_{j+1}^{\dagger}\hat{b}_{j+1}\hat{b}_{j}^{\dagger}\hat{b}_{j}\right],
\end{align}
\end{subequations}
with the different modes coupled locally through the particle non-conserving term
\begin{equation}
\hat{H}_{ab}=-\sum_{j=1}^{L}\varepsilon(\hat{a}_{j}\hat{b}_{j}+\hat{a}_{j}^{\dagger}\hat{b}_{j}^{\dagger}).
\end{equation}
Just as in the previous model, this term injects pairs of excitations in the system, but now the excitations are distinguishable (non-degenerate pair injection). The resulting Hamiltonian $\hat{H}=\hat{H}_{a}+\hat{H}_{b}+\hat{H}_{ab}$ is invariant under continuous transformations of the relative phase between the modes, that is, $\{\hat{a}_{j}\rightarrow e^{\mathrm{i}\theta}\hat{a}_{j},\hat{b}_{j}\rightarrow e^{-\mathrm{i}\theta}\hat{b}_{j}\}$ $\forall j$ and $\theta\in\mathbb{R}$, providing a $U(1)$ symmetry to the model.

We consider now a coherent-state ansatz of the form $\bigotimes_{j=1}^{L}|\alpha_{j},\beta_{j}\rangle$, with $\hat{a}_{j}|\alpha_{j},\beta_{j}\rangle=\alpha_{j}|\alpha_{j},\beta_{j}\rangle$ and $\hat{b}_{j}|\alpha_{j},\beta_{j}\rangle=\beta_{j}|\alpha_{j},\beta_{j}\rangle$, where $\{\alpha_{j}\in\mathbb{C},\beta_{j}\in\mathbb{C}\}_{j=1,2,...,L}$ are the variational parameters. The energy functional $\langle\hat{H}\rangle$ takes the form
\begin{align}
E=\sum_{j=1}^{L}\biggl[-\mu\left(\vert\alpha_{j}\vert^{2}+\vert\beta_{j}\vert^{2}\right) & -2J\text{Re}\left\{ \alpha_{j}\alpha_{j+1}^{*}+\beta_{j}\beta_{j+1}^{*}\right\} -2\varepsilon\text{Re}\left\{ \alpha_{j}\beta_{j}\right\}
\\
& +\frac{U}{2}\left(\vert\alpha_{j}\vert^{4}+\vert\beta_{j}\vert^{4}\right)+V\left(\vert\alpha_{j}\vert^{2}\vert\alpha_{j+1}\vert^{2}+\vert\beta_{j}\vert^{2}\vert\beta_{j+1}\vert^{2}\right)\biggr].\nonumber 
\end{align}
As in our original model, the $J$ term is minimized when the complex amplitudes of all sites have the same phase. In addition, the $\varepsilon$ term is minimized when each product $\alpha_{j}\beta_{j}$ is real and positive, that is, the phase sum of $\alpha_{j}$ and $\beta_{j}$ is an integer multiple of $2\pi$. Of course, the symmetry of the model allows the relative phase between $\alpha_{j}$ and $\beta_{j}$ to be arbitrary, that is, given a minimizing configuration $\{\alpha_{j},\beta_{j}\}_{j=1,2,...,L}$, the configuration $\{e^{\mathrm{i}\theta}\alpha_{j},e^{-\mathrm{i}\theta}\beta_{j}\}_{j=1,2,...,L}$ is also a minimum for any value of $\theta$. To simplify the derivation,
let us take $\theta=0$ as the representative minimum, specifically taking $\alpha_{j}$ and $\beta_{j}$ real and positive. The minimization condition reads now
\begin{subequations}\label{2mode-CoherentMinimumEquations}
\begin{align}
\frac{1}{2}\frac{\partial E}{\partial\alpha_{j}} & =\left(U\alpha_{j}^{2}-\mu\right)\alpha_{j}+V\alpha_{j}(\alpha_{j+1}^{2}+\alpha_{j-1}^{2})-J(\alpha_{j+1}+\alpha_{j-1})-\varepsilon\beta_{j}=0,
\\
\frac{1}{2}\frac{\partial E}{\partial\beta_{j}} & =\left(U\beta_{j}^{2}-\mu\right)\beta_{j}+V\beta_{j}(\beta_{j+1}^{2}+\beta_{j-1}^{2})-J(\beta_{j+1}+\beta_{j-1})-\varepsilon\alpha_{j}=0.
\end{align}
\end{subequations}
These equations possess ``balanced'' solutions with $\alpha_{j}=\beta_{j}$, as well as ``unbalanced'' ones with $\alpha_{j}\neq\beta_{j}$. The variational energy of the balanced ones is equal to twice the one we studied in Section \ref{sec:Appendix-Coherent-State} for the single-local-mode model, Eq. (\ref{CoherentEnergyFunctional}). Hence, such balanced configurations have the same energy landscape and phase diagram as the one we studied in the previous case. Specifically, a homogeneous solution dominates when $V/U\leq1/2+1/\nu$, with $\nu=(\mu+\varepsilon)/2J$, while a staggered solution appears when $V/U>1/2+1/\nu$. Hence, the coherent-state ansatz predicts a transition from a homogeneous superfluid to a supersolid phase in the balanced sector, in both cases breaking a continuous $U(1)$ symmetry spontaneously.

It is possible to find analytically or semi-analytically unbalanced solutions to (\ref{2mode-CoherentMinimumEquations}). For example, it is easy to see that homogeneous solutions exist only when $2\varepsilon\leq\mu+2J$, with
\begin{subequations}
\begin{align}
2\alpha_{j}^{2} & =\mu+2J\pm\sqrt{(\mu+2J)^{2}-4\varepsilon^{2}}\;\forall j,\\
2\beta_{j}^{2} & =\mu+2J\mp\sqrt{(\mu+2J)^{2}-4\varepsilon^{2}}\;\forall j.
\end{align}
\end{subequations}
These have an energy $E=-L[3\varepsilon^{2}+(\mu+2J)^{2}/2]/(U+2V)$, which is easily proven to be larger than that of the balanced homogeneous solutions, corresponding to twice that in Eq. (\ref{ESF}). It is possible to find staggered unbalanced solutions semi-analytically, but these are also shown to have larger energy than the balanced solutions in the region of parameters that we have discussed in this work. The same conclusion is drawn from the exhaustive search that we have carried by numerically simulating the imaginary-time dynamical equations equivalent to those derived in Section \ref{Sec:iTimeCoherent} for the previous model starting from random initial conditions,
\begin{subequations}
\begin{align}
\dot{\alpha}_{j} & =-\langle[\hat{a}_{j},\hat{H}]\rangle=\left[\mu-U\vert\alpha_{j}\vert^{2}+V\left(\vert\alpha_{j+1}\vert^{2}+\vert\alpha_{j-1}\vert^{2}\right)\right]\alpha_{j}+\varepsilon\beta_{j}^{*}+J(\alpha_{j+1}+\alpha_{j-1}),
\\
\dot{\beta}_{j} & =-\langle[\hat{b}_{j},\hat{H}]\rangle=\left[\mu-U\vert\beta_{j}\vert^{2}+V\left(\vert\beta_{j+1}\vert^{2}+\vert\beta_{j-1}\vert^{2}\right)\right]\beta_{j}+\varepsilon\alpha_{j}^{*}+J(\beta_{j+1}+\beta_{j-1}).
\end{align}
\end{subequations}

In summary, we have argued via a coherent-state ansatz that it is possible to extend the model studied in the main text in such a way that pair injection retains a full $U(1)$ symmetry while still leading to the same enhancement of supersolidity in the phase diagram.

\end{document}